\documentclass[useAMS,twocolumn,usenatbib]{mn2e}
\usepackage{amssymb}
\usepackage{amsmath}
\usepackage{graphicx}
\DeclareGraphicsExtensions{.eps, .ps, .eps.gz, ps.gz}
\usepackage[british]{babel} 
\usepackage{placeins}
\usepackage{subfigure}
\usepackage{hyperref}
\usepackage{color}

\newcommand{\dif}{{\mathrm d}}

\begin{document}
  

\title{Non-local bias contribution to third-order galaxy correlations}

\author[J. Bel, K. Hoffmann, E. Gazta\~naga]
{J. Bel$^{1}$, K. Hoffmann$^{2}$, E.Gazta\~naga$^{2}$\\
$^{1}$INAF - Osservatorio Astronomico di Brera, Via Brera 28, 20122 Milano, via E. Bianchi 46, 23807 Merate, Italy\\
$^{2}$Institut de Ci\`{e}ncies de l'Espai (ICE, IEEC/CSIC), E-08193 Bellaterra (Barcelona), Spain\\
} 

\date{Received date / Accepted date}

\maketitle


\begin{abstract}
  We study halo clustering bias with second- and third-order
  statistics of halo and matter density fields in the MICE Grand Challenge 
  simulation. We verify that two-point correlations deliver reliable estimates of the linear bias
   parameters at large scales, while estimations from the variance can be significantly affected by 
  non-linear and possibly non-local contributions to the bias function.
  Combining three-point auto- and cross-correlations we find, for the first time in configuration space,
  evidence for the presence of such non-local contributions.
  These contributions are consistent with predicted second-order non-local 
  effects on the bias functions originating from the dark matter tidal field.
  Samples of massive haloes show indications of bias (local or non-local) beyond second order.
  Ignoring non-local bias causes $20-30$\% and $5-10$\% overestimation of the linear bias 
  from three-point auto- and cross-correlations respectively.
  We study two third-order bias estimators which are not affected by second-order non-local 
  contributions. One is a combination of three-point auto- and 
  cross- correlation. The other is a combination of third-order one- and two-point cumulants.
  Both methods deliver accurate bias estimations of the linear bias. Furthermore 
  their estimations of second-order bias agree mutually. Ignoring non-local 
  bias causes higher values of the second-order bias from three-point correlations.
  Our results demonstrate that third-order statistics can be employed for breaking the
  growth-bias degeneracy.
  
\end{abstract}
 


\section{Introduction}\label{sec:introduction}

With the increasing amount of data coming from current and future large-scale galaxy 
surveys, errors on the observed statistical properties of the spatial galaxy distribution are rapidly
decreasing. This high level of precision requires at least the same level of accuracy in the
modelling of the corresponding observables.
An important observable is the growth of large-scale density fluctuations with time, which is 
sensitive to the universal matter density, the expansion of space as well as to 
the gravitational interaction of matter at large scales. Measurements of the this growth
therefore provide constrains on cosmological parameters
\citep[e.g.][]{rossetal07,C&G09,S&P09,SP&R12,reidetal12,delatorreetal13},
possible deviations from General Relativity \citep{GL, Lue04} or on alternative
phenomenological description for the accelerated expansion, such as the effective field theory
\citep{SB&M, PS&M14}.
Growth measurements can be undertaken by comparing the second-order correlations $\xi$ of
galaxy distributions at different redshifts. A critical aspect of this approach is the bias between the
correlations of galaxies and those of the full matter density field. This bias can either be predicted
with the peak-background split model \citep[e.g.][]{bardeen1986, cole1989, Sheth99},
or directly determined from observations using weak lensing observables, redshift space distortions
or reduced third-order correlations at large scales. The latter method relies on the fact that such
third-order correlations are independent of the growth
at large scales, but sensitive to the bias. Third-order galaxy correlations therefore have the potential
to tighten constrains on cosmological models from observations \citep[e.g.][]{marin11, marin13}.
However, how useful third-order correlation are for this purpose depends on the accuracy and
the precision of the bias estimations they deliver \citep[e.g.][]{Wu2010, Eriksen2015}.

The present work is part of a series of papers with which we aim to characterise and understand
differences between various bias estimations using the huge volume of the MICE Grand challenge
simulation. We thereby assume that galaxies can be associated with dark matter haloes.
This analysis includes bias measurements derived from a direct comparison between
halo and matter fluctuations (Bel, Hoffmann, Gazta\~naga, in preparation) as well as bias predictions
from the peak-background split model \citep{paperIII}. Continuing this series we now aim at
deepening our previous work on the bias from third-order correlations \citep{paperI}.

In this previous work we investigated growth measurements based on bias estimations from
third-order halo auto-correlations (halo-halo-halo). Despite the positive and encouraging
outcome of this analysis, we pointed out some discrepancies between the first- and
second-order bias estimations derived from two different third-order statistics. The 
first third-order statistics is the reduced three-point correlation function $Q$ \citep{Fry84a}.
The second one is based on a particular combination \citep{Szapudi98, bm} of the skewness $S_3$
and the correlator $C_{12}$ \citep{bernardeau02} \citep[referred to as $\tau$ method,][]{bm},
which correspond to the three-point correlation for triangles with one and two collapsed legs
respectively. We found that linear bias estimations from the three-point auto correlations
over-estimate the true linear bias (probed by two-point correlations at large scales)
by $20$-$30$\%. The $\tau$ method delivers accurate linear bias estimations, but with lower precision.
Furthermore, its bias estimation become unreliable when samples of massive haloes
($M>10^{14}h^{-1}M_\odot$) are considered. Understanding these discrepancies between different
bias estimators is crucial for constraining cosmological models with observed third-order galaxy statistics.
In the light of previous studies \citep{M&G11, Pollack12, chan12, baldauf12} we pointed 
out that such discrepancies might originate from non-linear and/or non-local effects on
the different bias estimators. Hence, in the present paper, we focus on the analysis of third-order
cross-correlations (halo-matter-matter) to decrease the impact of non-linearities on our bias
estimators in order to investigate the extension of the local bias model to a possible non-local
component.

We recall the bias expansion, introduced by \citet{FG}, which assumes a local relation between the
density contrast of matter and haloes

\begin{equation} 
\delta_h=F[\delta_{m}] \simeq \sum_{i=0}^{N}\frac{b_i}{i!}\delta_{m}^i.
\label{eq:biasfunction}
\end{equation}
In the above relation we assume that $\delta_h$ and $\delta_m$
  are coarse grain quantities obtained by applying a smoothing
  process \citep[see][for a discusion of the effect of the smoothing]{M&G11, B&K99}.
Inaccuracies of this deterministic relation (equation \ref{eq:biasfunction}) might arise from tidal forces in the matter field, leading
to a non-local contribution in the biasing relation. At second order it can be expressed as

\begin{equation}
\delta_h({\bf x}) = b_1 \left\lbrace \delta_{m}({\bf x}) + \frac{c_2}{2}(\delta_{m}^2({\bf x}) - \langle \delta_{m}^2 \rangle) + \frac{\gamma_2}{b_1} \mathcal{ G}_2({\bf x}) \right\rbrace,
\label{eq:def_biasmodel}
\end{equation}
where $\gamma_2$ represents the non-local bias parameter \citep{chan12, baldauf12}.
This non-local component depends on the divergence $\theta_v$ of the normalised 
velocity field (${\bf v}/\mathcal{H}/f$)

\begin{equation}
\mathcal{ G}_2({\bf x})=- \int \beta_{12}\theta_v({\bf q}_1) \theta_v({\bf q}_2) \hat W[q_{12}R]e^{i {\bf q}_{12}\cdot {\bf x} }d^3 {\bf q}_1 d^3 {\bf q}_2,
\label{eq:g2}
\end{equation}
where $\beta_{12}\equiv 1 - \left( \frac{ {\bf q}_1\cdot {\bf q}_2 }{q_1q_2} \right )^2$ represents
the mode-coupling between density oscillations with wave vectors ${\bf q}$, 
which describe tidal forces. $W[q_{12}R]$ is the Fourier transform of a spherical Top-hat window
with radius $R$.
In order to be consistent with the definition of second-order
bias parameter, in this paper we shall refer to the non-local component of the biasing relation
(\ref{eq:def_biasmodel}) using the quantity $g_2\equiv 2\frac{\gamma_2}{b_1}$.

Evidence for significant contributions of such a non-local component to bias function has been 
reported in Fourier space for different simulations \citep{chan12, baldauf12}. 
However, it remains unclear how strongly these non-local contributions affect the bias and
consequently third-order statistics of large-scale halo distributions in configuration 
space. We address this latter question in the present study and suggest possibilities to employ
third-order statistics for accurate bias measurements, independently of non-local bias.

This paper is organised as follows. In Section \ref{sec:simulation} we briefly present the simulations
on which our analysis is based on, in Section \ref{sec:bias_estimators} we present the
bias estimators implemented in this paper and in Section \ref{sec:results} we discuss and comment
our results. Finally we summarise our work and draw conclusions in Section \ref{sec:conclusion}.
In the Appendix \ref{sec:nloc_bias} we present a perturbative approach
to describe the effect of non-local bias on third-order statistics in configuration space.
\section{Simulation}\label{sec:simulation}
    
      Our analysis is based on the Grand Challenge run of the Marenostrum
      Institut de Ci\`encies de l'Espai (MICE) simulation suite to which we refer to
      as MICE-GC in the following.
      Starting from small initial density fluctuations at redshift $z=100$, the formation of large scale
      cosmic structure was computed with $4096^3$ gravitationally interacting collisionless particles
      in a $3072$ $h^{-1}$Mpc box using the GADGET - 2 code \citep{springel05} with a softening
      length of $50$ $h^{-1}$kpc. The initial velocities and particle displacements were generated using the Zel'dovich
      approximation and a CAMB power spectrum with the power law index of $n_s = 0.95$, 
      wich as normalised to fulfil $\sigma_8 = 0.8$ at $z=0.0$. The cosmic expansion is described by the
      $\Lambda$CDM model for a flat universe with a mass density of
      $\Omega_m$ = $\Omega_{dm} + \Omega_b = 0.25$. The density of the
      baryonic mass is set to $\Omega_b = 0.044$ and $\Omega_{dm}$ is the dark matter density.
      The dimensionless Hubble parameter is set to $h = 0.7$. More details and validation test on this
      simulation can be found in \citet{mice1}, \citet{mice3} and \citet{paperI}.
      
      Dark matter haloes were identified as Friends-of-Friends groups \citep{davis85} with a
      redshift independent linking length of $0.2$ in units of the mean particle separation.
      These halo catalogs and the corresponding validation checks are presented in \citet{mice2}.
      
      As in our previous analysis \citep[][]{paperI} we divide the haloes into the four redshift
      independent mass samples M0, M1, M2 and M3, specified in Table \ref{table:halo_masses}.
      These samples span a mass range from Milky Way like haloes up to massive galaxy clusters.
      \begin{table}
      \centering
        \caption{Halo mass samples. $N_p$ is the number of dark matter particles per halo, $N_{halo}$ is the
        number of haloes per sample in the comoving output at redshift $z=0.5$.}
        \label{table:halo_masses}
        \begin{tabular}{c  c c c }
       sample  & mass range [$10^{12} h^{-1}M_{\odot}$]   &  $N_p$ & $N_{halo}$\\
           \hline 
             M0	&	$0.58 - 2.32$	&	$20-80$	    &  $122300728$ \\
             M1	&	$2.32 - 9.26$	&	$80-316$	    &  $31765907$ \\
             M2	&	$9.26 - 100$	&	$316-3416$   &  $8505326$ \\
             M3	&	$\ge 100$	      &	$\ge 3416$	    &  $280837$ \\
            \hline
         \end{tabular}
         \label{table:halo_masses}
      \end{table}
      %
      For studying the bias estimators we use dark matter particles and haloes identified in the
      comoving outputs at redshift $z = 0.0$ and $0.5$.
\begin{figure}
\centerline{ \includegraphics[width=80mm,angle=0]{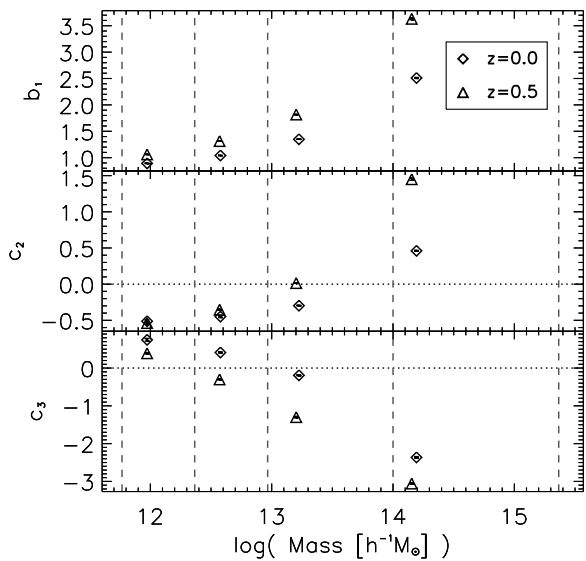}}
	\caption{
	Peak-background split prediction of the bias parameters $b_1$, $c_2$ and $c_3$
	for the mass samples M0-M3 versus the mean halo mass of each sample at redshift
	$z=0.0$ and $z=0.5$ (diamonds and triangles respectively). The mass ranges of the samples are 
	marked by vertical dashed lines.}
	\label{fig:pbs_bias}
\end{figure}

\subsection{Bias predictions}
The peak-background split model predicts that the bias parameters in
equation (\ref{eq:biasfunction}) are a function of the halo mass
\citep{bardeen1986, cole1989, Sheth99}. For deriving these predictions we fit the
mass function with the model of \citet{Tinker10}. The low mass sample M1 is thereby
excluded from the fitting range, since we expect the mass function measurements to be
affected by noise in the halo detection in this mass range
\citep[see][for a detailed discussion]{paperIII}. However, this exclusion does not
affect the sign of the predicted bias coefficients. The mean bias coefficients of each mass sample,
shown in Fig. is then obtained by weighting the bias prediction with the normalised halo
mass distribution.


In Fig. \ref{fig:pbs_bias} we show the predicted mass dependency for bias coefficients up to third order.
In \citet{paperIII} we compare the predictions for the linear bias parameter with 
measurements from two-point cross-correlations and find an agreement at the 
$10$\% level. Comparing the second- and third-order bias predictions to 
measurements from third-order statistics and a direct analysis of matter and 
halo density fluctuations we find a good qualitative agreement (Bel, Hoffmann, Gazta\~naga,
in preparation).
We shall use these predictions for discussing our results in Section \ref{sec:bias_estimators}.


\section{Bias estimators}\label{sec:bias_estimators}
	
	In this section we study various bias estimators from second- and third-order
	clustering statistics of haloes and matter, in order to quantify
	and understand differences between these estimations. Such an understanding is crucial for using
	third-order statistics in order to break the degeneracy between the linear galaxy (or halo)
	bias and the linear growth of matter fluctuations, as we discussed in 
	\citep{paperI}.
	
	In this previous study we found that the linear bias
	from the reduced three-point correlation in configuration space tends to overestimate
	the linear bias from the two-point correlation, even when the analysis is performed
	at very large scales ($>30 \ h^{-1}$Mpc). Similar findings have been reported in the
	literature \citep[e.g.][]{M&G11}. Such deviations can be expected from non-local
	contributions to the bias function \citep{chan12, baldauf12}. Furthermore, non-linear
	terms in the perturbative expansion of correlations functions, which are usually neglected
	in the analysis of clustering measurements, can contribute to the deviations between
	the different bias estimators \citep{Pollack12}. The goal of this section is to investigate
	the effect produced on various bias estimators of non-linear
	(sub-section \ref{sec:second_order_bias}) and non-local
	(sub-sections \ref{sec:3pc_bias} and \ref{sec:btau}) contributions to the biasing function. 
      
      Throughout this investigation we shall use the following notations: the density
      contrast of a stochastic field at position ${\bf r_i}$ will be shortly written as
      $\delta_{y,i}\equiv \delta_y({\bf r_i} )$ where $y$ refers to either haloes ($h$) or matter ($m$).
      Note that, when no confusion is possible, we will drop the index $i$ referring to the position. 

	\subsection{Bias from second-order statistics}\label{sec:second_order_bias}

      We study second-order statistics of the halo- and matter density fields in 
      terms of the two-point correlation and the variance. The density fields
      are therefore smoothed with a spherical Top-hat window function of radius 
      $R$. The two-point correlations can be defined as the average products of density
      contrasts at the positions $\bf{r}_1$ and $\bf{r}_2$,
      	\begin{equation}
		\xi^{xy}  \equiv \langle  \delta_{x,1} \delta_{y,2} \rangle (r_{12}),
		\label{eq:def_2pc}
	\end{equation}
	which is a function of the distance $r_{12}=|\bf r_2 - \bf r_1|$. If the two density 
	contrasts belong to the same density fields (haloes or matter, i.e. $x = y$) equation
	(\ref{eq:def_2pc}) defines the two-point auto-correlation, denoted in the following as $\xi$ 
      and $\xi_h$ for the matter and the halo field respectively. Note that,
      for a given smoothing window, the variance $\sigma^2$
      corresponds to the two-point correlation in the limit where $r_{12}=0$. In the case of two different 
      density fields (halo and matter) equation (\ref{eq:def_2pc}) defines the two-point cross-correlation,
      which we denote as $\xi^\times$ in the following.
      
      The average $\langle \ldots \rangle$ refers to all pairs, independently of their orientation. 
	We thereby follow the common assumption of ergodicity such that the ensemble average
      can be estimated from the mean over the comoving volume in each simulation snapshot.
	The two-point correlation is therefore an isotropic quantity, in contrast to 
	the three-point correlation, which we study later in Subsection \ref{sec:3pc_bias}.
	 Note that we ignore redshift space distortions in this study.

      If the biasing relation between matter and haloes was linear ($\delta_h=b_1\delta_m$) then
      the halo variance $\sigma_{h}^2\equiv\langle\delta_h^2\rangle$,
      the halo two-point auto-correlation $\xi_{h}\equiv\langle\delta_{h,1}\delta_{h,2}\rangle$,
      the halo-matter cross-variance $\sigma_{\times}^2\equiv\langle\delta_{h}\delta_{m}\rangle$
      and the halo-matter two-point cross-correlation
      $\xi^{\times}\equiv\langle\delta_{h,1}\delta_{m,2}\rangle$, 
      would provide equivalent estimators of the bias parameter $b$. Hence, we can define four
      linear bias estimators based on second-order statistics as
        
	\begin{eqnarray}
      b_{\sigma} & \equiv & \sigma_{h}/\sigma \label{eq:bsig}, \\
	b_{\sigma}^\times & \equiv & \sigma_\times^2/\sigma^2 \label{eq:bsigcross}, \\
	b_{\xi} & \equiv &  \sqrt{\xi_h/\xi} \label{eq:bxi}, \\
	b_{\xi}^\times & \equiv  &  \xi^\times/\xi \label{eq:bxicross}.
	\end{eqnarray}
      Any deviation from the equivalence  of these estimators provides a way to study the impact of possible non-linear 
      terms in the biasing relation between matter and haloes
      \citep[e.g.][respectively]{MS&S10, Saito14}.
      In fact, if the biasing function is a local relation, defined by its Taylor expansion
      (see equation (\ref{eq:biasfunction})), one can approximate the auto- and cross-correlation
      of haloes as a function of one- and two-point statistics of the matter field. \citet{FG} and \citet{bm}
      provide the corresponding expressions respectively for one- and two-point cumulants up to
      fifth order (i.e. $\langle\delta_{h,1}^n\delta_{h,2}^m\rangle$ and $n+m \le 
      5$), keeping terms up to second order (leading order plus one) in terms of $\xi$ and $\sigma^2$.

      Since we focus on non-linear effects, we express the second-order
      statistics of haloes (auto and cross) up to two orders after the leading order, which is keeping
      only terms of third-order in $\xi$ and $\sigma^2$ and lower. In the case of the two-point
      auto-correlation we obtain
      
      \begin{eqnarray}
      \frac{\xi_h}{b_1^2} & \simeq & \xi + (c_2C_{12}+c_3)\xi\sigma^2 + \frac{c_2^2}{2}\xi^2  \nonumber \\
       &     & + \left ( \frac{c_3}{3}C_{13} + c_4(\frac{C_{12}}{2}+ \frac{S_3}{3}) + \frac{c_5}{4} \right. \nonumber \\
       &     & \left. + \frac{c_2^2}{4}C_{22} + \frac{c_2}{2}c_3C_{12} + \frac{c_3^2}{4} \right )\xi\sigma^4 \nonumber \\
       &     & + \left ( c_2c_3C_{12} + \frac{c_2}{2}c_4 \right)\xi^2\sigma^2 + 
       \frac{c_3^2}{6}\xi^3,
       \label{eq:xiautoexpand}
      \end{eqnarray}
      while we find for the cross-correlation
      
      \begin{eqnarray}
      \frac{\xi^\times}{b_1} & \simeq & \xi + \frac{1}{2}(c_2C_{12}+c_3)\xi\sigma^2  \nonumber \\
      &     & + \frac{1}{2}\left ( \frac{c_3}{3}C_{13} + c_4(\frac{C_{12}}{2}+ \frac{S_3}{3})  +  \frac{c_5}{4} \right )\xi\sigma^4,
      \label{eq:xicrossexpand}
      \end{eqnarray}
      where $C_{nk}$ and $S_{n}$ are respectively the reduced correlators and
      cumulants of the matter field, defined as
      $C_{nk}\equiv \langle\delta_{m,1}^n\delta_{m,2}^k\rangle_c/\xi/\sigma^{2(n+k-2)}$ and
      $S_{n}\equiv \langle\delta_m^n\rangle_c/\sigma^{2(n-1)}$ \citep[see][]{bernardeau96}.
      The non-linear terms in $\xi$, present in equation (\ref{eq:xiautoexpand}) and
      (\ref{eq:xicrossexpand}), differ from each other. This means that non-linear bias affects
      the auto- and cross-correlation function of haloes differently. 
      The corresponding expansions
      for the variance and the cross-variance of haloes can be straightforwardly obtained by taking
      the one-point limit in the equations (\ref{eq:xiautoexpand}) and  (\ref{eq:xicrossexpand})
      (i.e. $r_{12}=0$). Hence, the reduced correlator $C_{nk}$ corresponds to the reduced cumulant
      $S_{n+k}$ and the two-point correlation function (auto and cross) converges to the variance
      (auto and cross).
      These two expansions show that, at third order in $\sigma^2$ and $\xi$, one
      needs to take into account bias parameters up to fifth order ($c_5$).  However, since
      we focus our analysis on scales at which the matter density contrast is small ($\delta_m \ll 1$),
      we will approximate the bias function as a third-order polynomial.
      In the following, we therefore set the fourth- and fifth-order parameters to zero.

      The equations (\ref{eq:xiautoexpand}) and (\ref{eq:xicrossexpand}) show that, if only the leading
      order terms are kept, then the bias estimators (\ref{eq:bsig})-(\ref{eq:bxicross}) converge to the linear
      bias $b_1$. However, they also tell that, if higher order terms are contributing to the 
      signal, then significant deviations between the estimators must be detectable. In the following,
      we exploit this fact to test the impact of non-linearities on the
      estimators (\ref{eq:bsig})-(\ref{eq:bxicross}).

      The auto- and cross-correlation functions are estimated by applying the estimator proposed by 
      \citet{bm} on the halo and the dark matter particle 
      distributions. Its implementation in the MICE-GC simulation is described in \citet{paperI},
      to estimate the errors we use a Jack-knife method on $64$ cubical cells.
      For studying the scale dependence we vary the smoothing scale $R$, while we keep the
      correlation length ratio $r_{12}/R = 2$ fixed.
      %
      \begin{figure}
      \includegraphics[width=85mm,angle=0]{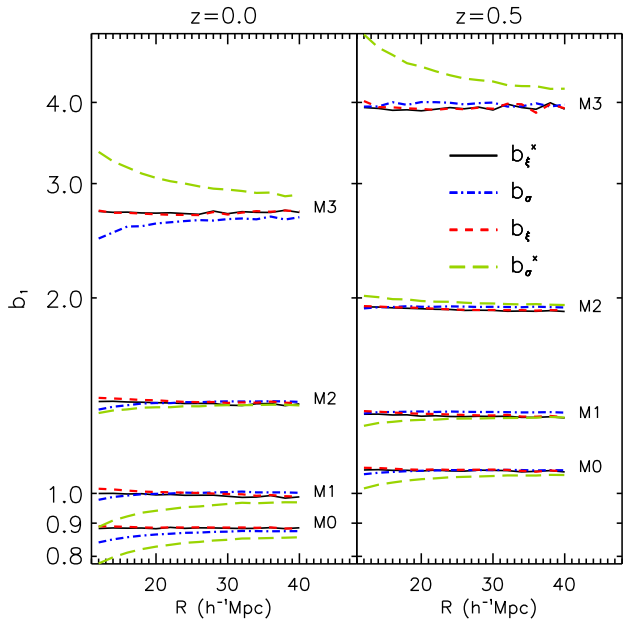}
      \caption{Comparison between four linear bias estimations from measurements
      of the auto- and cross- variance and the two-point auto- and cross- correlation
      (blue, green, red and black respectively), derived via equation (\ref{eq:bsig}) - 
      (\ref{eq:bxicross}). Results for the redshifts $z=0.0$ and $z=0.5$ are shown in the
      left and right panel respectively. The different smoothing scales $R$ of the halo and
      matter density fields, used for the bias estimations, are shown on the $x$-axis.
      For measuring $\xi$ and $\xi^\times$ we fix $r_{12} = 2R$.}
      \label{fig:1order_bias_comp}
      \end{figure}
            
      The bias measurements for the four mass samples M0-M3 at the redshifts
      $z=0.0$ and $z=0.5$, derived from the estimators (\ref{eq:bsig})-(\ref{eq:bxicross})
      are presented in Fig. \ref{fig:1order_bias_comp}. We can see that, on the
      considered scale range, the estimators involving two-point statistics are in good agreement with
      each other and display negligible scale dependence. This indicates that higher-order terms in the
      equations (\ref{eq:xiautoexpand}) and (\ref{eq:xicrossexpand}) can indeed be safely neglected.
      However, estimators involving the variance differ significantly from $b_\xi$ and $b_\xi^\times$,
      especially at small scales when the matter variance $\sigma^2$ approaches unity. Furthermore, they
      do not mutually agree, even on large scales.  Similar results have been obtained by \citet{MS&S10}
      and \citet{M&G11}. In order to explain differences between $b_{\xi}$ and $b_{\xi}^\times$
      \citet{MS&S10} set $c_3=0$ and drop the term in $\sigma^4$ into the higher order expansion;
      although they did not have a theoretical motivation for that.
     
	Following their approach and using the equations (\ref{eq:xiautoexpand}) and (\ref{eq:xicrossexpand})
	we expand the expressions (\ref{eq:bsig}) - (\ref{eq:bxicross}) up to the second-order in
	$\sigma^2$ and $\xi$ and express $b_\sigma$, $b_\xi$ and $b_\sigma^\times$ with respect to $b_\xi^\times$
	
	\begin{eqnarray}
	\left (\frac{b_{\sigma} }{b_1}\right )^2 -  \left (\frac{b_{\xi}^\times }{b_1}\right )^2\!\!\!  &\!\!\!  = \!\!\! & \!\!\!  \left[ c_2(S_3-C_{12}) + \frac{c_2^2}{2} \right] \sigma^2 \nonumber \\
	& & \!\!\! + \left[   \frac{c_2^2}{4}(S_4-C_{12}^2) + \frac{c_3}{3}(S_4-C_{13}) \right. \nonumber \\
        & & \!\!\! + \left. \frac{c_2c_3}{2}(3S_3-C_{12}) + \frac{c_3^2}{6}  \right] \sigma^4
        \label{one} \\
	\left (\frac{b_{\sigma}^\times }{b_1}\right )^2 \!\!\! -  \left (\frac{b_{\xi}^\times }{b_1}\right )^2 \!\!\! & \!\!\! =\!\!\! & \!\!\!  c_2(S_3-C_{12}) \sigma^2 + \left[ \frac{c_2^2}{4}(S_3^2-C_{12}^2) \right. \nonumber \\
	& &\!\!\!  \left. + \frac{c_3}{3}(S_4-C_{13})  \right . \nonumber \\
	& &\!\!\!  \left. + \frac{c_2c_3}{2}(S_3-C_{12}) \right] \sigma^4
	\label{two} \\
	\left (\frac{b_{\xi} }{b_1}\right )^2\!\!\!  -  \left (\frac{b_{\xi}^\times }{b_1}\right )^2\!\!\!  &\!\!\!  =\!\!\!  &\!\!\!  \frac{c_2^2}{2} \xi +  \frac{c_2^2}{4}\left[ C_{22}-C_{12}^2  \right] \sigma^4 \nonumber \\
	& &\!\!\!  +  c_2c_3C_{12}\xi\sigma^2 + \frac{c_3^2}{6}\xi^2.
	\label{three}
	\end{eqnarray}
	With the expressions above one can partially explain the deviations between the 
	different bias estimators, displayed in Fig. \ref{fig:1order_bias_comp}.  We thereby rely on
	the fact that differences between the linear bias and differences between the 
	second power of these values are monotonically related to each-other, due to 
	the positive amplitude of the linear bias.
      The first interesting point is to notice that equation (\ref{three}) does not exhibit any contribution
	from $\sigma^2$ at the leading order.
	Moreover the contribution from $\sigma^4$ vanishes since
	it has been shown that correlators obey the factorisation rule $C_{nk}=C_{1n}C_{1k}$ 
      \citep{bernardeau96}. As a result equation  (\ref{three}) contains only terms in $\xi$, which is small on
	the considered scale range ($r_{12} \gtrsim 20h^{-1}$Mpc). This explains the good agreement between $b_\xi$ and $b_\xi^\times$
	for all halo mass ranges. 
	
	The expression which we obtained for $b_\xi$ differs from the one given by \citet{MS&S10}
	 presumably due to a truncation of the higher-order expansion of the halo two-point
	cross-correlation function by these authors. In addition, our results seem to confirm the
	necessity they had to set the contribution of order $\sigma^4$ and the third-order bias $c_3$
	to zero in order to explain their results.
	Note also that at very small scales, Fig. \ref{fig:1order_bias_comp} shows
	that $b_\xi$ tends to be larger than $b_\xi^\times$. This small tendency can be explained by
	the fact that the leading order term of equation (\ref{three}) is positive.
	
	Regarding the comparison between $b_\sigma^\times$ and $b_\xi^\times$, the leading order
	term in expression (\ref{two}) shows that their relative amplitude
	depends on the sign of $c_2$. If it is negative $b_\sigma^\times$ is lower than 
	$b_{\xi}^\times$, which is the case for the mass samples M0-M2 at $z=0$ and M0-M1 at 
	$z=0.5$. The peak-background split bias predictions, displayed in Fig. \ref{fig:pbs_bias}, shows that
	$c_2$ is indeed expected to be negative for these mass samples (see Section \ref{sec:simulation}
	for details about the prediction). The deviations between the
	bias from the variance and the two-point cross-correlation is, at low masses, in good agreement
	with the one predicted by equation (\ref{one}).
	In fact, the leading order term contains two contributions, one in $c_2^2$
	and another in $c_2$. The latter being dominant for low masses where $|c_2|<1$. It
	follows that $b_\sigma$ is expected to lie between $b_\sigma^\times$ and $b_\xi^\times$ which
	is confirmed by Fig. \ref{fig:1order_bias_comp}.  However, for higher mass bins we observe
	an opposite tendency to the expected one. By combining equations (\ref{one}) and (\ref{two})
	one obtains that the relative position between $b_\sigma$ and $b_\sigma^\times$ is
	
	\begin{eqnarray}
	\left (\frac{b_{\sigma} }{b_1}\right )^2 \!\!\!- \!\!\! \left (\frac{b_{\sigma}^\times }{b_1}\right )^2 \!\!\!& \!\!\!= \!\!\!&  \!\!\! \frac{c_2}{2}\sigma^2 \!\! +\!\! \left[  c_2c_3S_3+\frac{c_2^2}{4}(S_4\!\!-\!\! S_3^2) + \frac{c_3^2}{6} \right]\!\! \sigma^4.  \label{four}
	\end{eqnarray}
	As a result when $c_2 > 0$, the bias from the variance is expected to be greater than the bias
	from the cross-variance. Note also that, even in the particular case of $c_2\simeq 0$ (M2 at $z=0.5$),
	the remaining term in $c_3^2$ shows that $b_\sigma$ must be greater than $b_\sigma^\times$.
	The measurements, shown in Fig. \ref{fig:1order_bias_comp}, exhibit a different behaviour than
	the one just described.
	In fact, for the high mass bins M3 at $z=0$ and M2, M3 at $z=0.5$, the estimate from
	$b_\sigma$ stays below the one from $b_\sigma^\times$. The bias from the variance
	apparently saturates to the value of the bias from two-point correlations and therefore seems to 
	constitute a more robust linear bias estimator than the cross-variance. Note that, this tendency
	is observed even at very large scale $R=40 \ h^{-1}$Mpc which suggests that non-linear corrections
	are not responsible. 

       We conclude that higher-order contributions arising from non-linear bias can only explain partially
       the relative amplitudes between the linear bias estimators displayed in Fig.~\ref{fig:1order_bias_comp}.
       This result motivates the introduction of a non-local component in the biasing relation, as suggested in
      the literature \citep[e.g.][]{chan12, baldauf12, Saito14} and also shows that one needs
      to go beyond second-order statistics. However, we do not expect non-local bias to
      significantly affect our  second-order halo statistics,
      especially because of the large smoothing scales
      applied in this analysis. This might not be the case for third-order statistics.
       We therefore speculated in \citet{paperI} that a non-local component in the biasing relation
       could explain the differences between second- and third-order bias estimators. In the next
       subsection, we therefore generalise our analysis in studying linear and quadratics bias estimation
       from third-order statistics (auto and cross). Finally, we note that the bias estimator
       based the two-point cross-correlation function ($b_\xi^\times$) can be considered as a reliable
       estimator of the linear bias $b_1$, we thus consider it as a reference in the following.

	\subsection{Bias from three-point cross-correlation $b_Q^{\times}$}\label{sec:3pc_bias}
	
	The three-point correlation of the three generic density
        fields that we label $x$, $y$ and $z$ at the 
	positions ${\bf r_1}$, ${\bf r_2}$ and ${\bf r_3}$ (respectively) is defined as
	
	\begin{equation}
		\zeta^{xyz}  \equiv \langle  \delta_{x,1} \delta_{y,2} \delta_{z,3} \rangle,
		\label{eq:def_3pc}
	\end{equation}
	where the vectors ${\bf r}_{1}$,  ${\bf r}_{2}$ and ${\bf r}_{3}$ form a closed triangle which can
	be parametrised in terms of the size of its three legs $r_{ij}\equiv |{\bf r_j} - {\bf r_i}|$ or in
	terms of the two legs ${\bf r}_{12}$, ${\bf r}_{13}$ and the angle
	$\alpha_{23} = \text{acos}({\bf \hat r}_{12} \cdot {\bf \hat r}_{13})$ between them.
	 As for the two-point correlation in Section \ref{sec:second_order_bias} $\langle \ldots \rangle$
	denotes the mean over all triangles in the analysed volume. However, in 
	contrast to the two-point correlation, where the average is taken over all pairs of $\delta$
	independently of their direction, $\zeta$ is not isotropic as it is sensitive 
	to the shape of the large-scale structure. It therefore provides access to additional 
	information.
	In case of the auto-correlation the three density contrasts in equation (\ref{eq:def_3pc})
	refer to the same density field. The corresponding three-point correlation function 
	of haloes and matter are therefore $\zeta^{hhh}$ and $\zeta^{mmm}$, respectively. To measure
	the bias from the three-point cross-correlation between haloes and matter densities we
	compute $\zeta^{hmm}$, $\zeta^{mhm}$, $\zeta^{mmh}$. These quantities
	are then compared to the hierarchical three-point cross-correlation \citep{Fry84b}
      
      \begin{equation}
      		\zeta^{hm} _H \equiv \xi^{hm}_{12} \xi^{hm}_{13}  + \xi^{mh}_{12} \xi^{hm}_{23}  + \xi^{mh}_{13} 
      		\xi^{mh}_{23}.
      		\label{eq:def_3pcH}
      \end{equation}
       Note that here $\xi^{hm}_{ij}$ referrers to the two-point cross-correlation between haloes at
       position ${\bf r_i}$ and matter at position ${\bf r_j}$, which is called $\xi^\times$ in the
       remainder of this article.
      Combining equation (\ref{eq:def_3pc}) and (\ref{eq:def_3pcH}) one can define
	the symmetric reduced three-point cross-correlation function \citep{Pollack12}

	\begin{equation}
          	Q_h^{\times}\equiv \frac{1}{3}\frac{\zeta^{hmm}  + \zeta^{mhm} + \zeta^{mmh}}{\zeta^{hm} _H}.
          	\label{eq:def_Q3cross}
        \end{equation}
      The reduced three-point auto-correlations for matter and halo density 
      fields are defined analogously as $Q_m \equiv \zeta^{mmm} /\zeta_H^{mm}  $ and $Q_h \equiv 
      \zeta^{hhh}/\zeta_H^{hh}$. This way $Q_h^\times$, $Q_h$ and $Q_m$ quantify
      any departure from the hierarchical ansatz \citep{Fry84a}. In the following we will refer to the
      reduced three-point correlation as the three-point correlation. 

	\begin{figure}
	   \centering
	   \includegraphics[width=80mm, angle = 270]{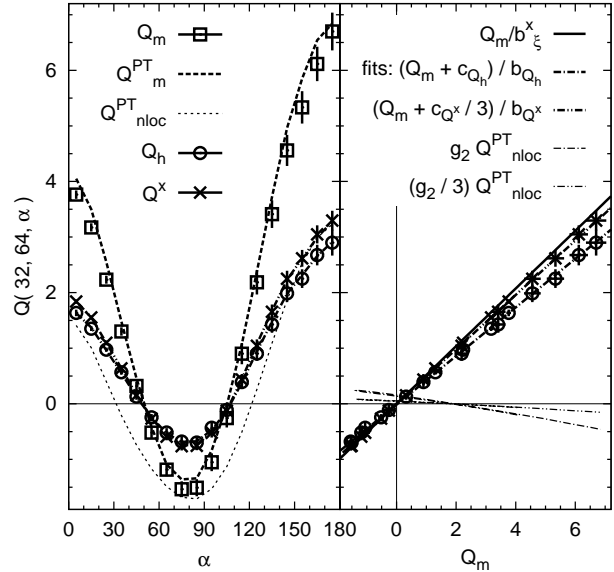}
	   \caption{
	       {\it Left}: three-point auto- and cross-correlation for matter and haloes
  	         ($Q_h$, $Q^{\times}$ and $Q_m$) in the mass sample M2 at redshift $z=0.5$, measured
  	         using triangles  with fixed legs of $32$ and $64$ $h^{-1}$Mpc for different opening angles
		    $\alpha$ (circles, crosses and squares respectively).
		    Fits based on the local bias model (i.e. $g_2=0$) to $Q_h$ and $Q^{\times}$
		    from equation (\ref{eq:b1c2_q3auto}) and (\ref{eq:b1c2_q3cross})
		    are shown as thick dashed-dotted and dash-double dotted lines.
		     Predictions for the matter three-point correlation and the non-local
		    component from perturbation theory ($Q^{PT}_m$ and $Q^{PT}_{nloc}$) are shown
		    as thick and thin dashed lines respectively.
		    {\it Right}: the $Q_h$ and $Q^{\times}$ versus $Q_m$ relation, used for deriving the
		    linear and quadratic parameters $b_1$ and $c_2$ in the local bias model.
		     Thin dashed-dotted and dash-double dotted lines show the non-local
		    contributions to $Q_h$ and $Q^{\times}$ respectively, using $g_2$ from
		    $\Delta Q_{cg}$, equation (\ref{eq:deltaqcg}).}
	   \label{fig:q3cross_dmvshalo}
	\end{figure}

      Inserting the non-local quadratic bias model (equation (\ref{eq:def_biasmodel})) into the definition of the
      three-point correlation for haloes yields, via a second-order perturbative expansion, in the limit
      of small density fluctuations and large triangles
      
	\begin{equation}
		Q_h =\frac{1}{b_1}\bigl\{ Q_m+ [c_2 + g_2Q_{nloc}]\bigr\},
		\label{eq:b1c2_q3auto}
	\end{equation}
	which can be generalised to the case of three-point cross-correlation,

	\begin{equation}
		Q_{h}^\times = \frac{1}{b_1}\bigl\{ Q_m+ \frac{1}{3}[c_2 + g_2Q_{nloc}]\bigr\}
		\label{eq:b1c2_q3cross}
	\end{equation}
       (see Appendix \ref{sec:nloc_bias}). These expressions differ significantly from the ones obtained from the local bias model,
      as they include the non-local contribution to the three-point halo correlation 
      $Q_{nloc}$, which we present in more detail below. The expression for the local model, which we
       assumed in \citet{paperI} corresponds to a vanishing non-local bias parameter, i.e. $g_2=0$.
	Since $Q_{nloc}$ is a function of the opening angle $\alpha\equiv\alpha_{23}$, the
	three-point halo auto- and halo-matter cross-correlations are therefore no longer
	linearly related to the matter three-point correlation. This $\alpha$ dependence arises from
	the fact that $Q_{nloc}$ originates from tidal forces, which modify the shape of matter fluctuations.
      %

	\begin{figure}
	   \centering
	   \includegraphics[width=70mm, angle = 270]{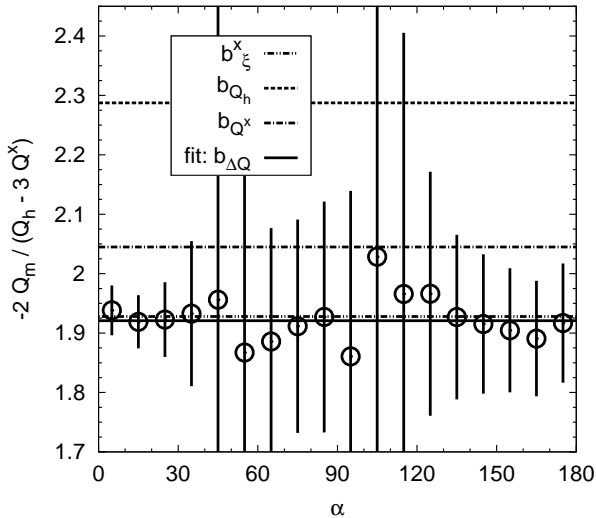}
	   \caption{
		    Linear bias of the halo mass sample M2 at the redshift $z=0.5$, measured
		    independently of second-order contributions (local and non-local) to the three-point
		    correlation via equation (\ref{eq:bdQ}) by combining $Q_h$, $Q^{\times}$ and $Q_m$
		    (open circles). The fit to the measurements is shown
		    as solid line. The linear bias measurements, derived from
		    $\xi^\times$, $Q_h$ and $Q^{\times}$ within the local bias model (via equation
		    (\ref{eq:bxicross}), (\ref{eq:b1c2_q3auto}) and (\ref{eq:b1c2_q3cross})
		    respectively with $g_2=0$) are shown as dashed, dash-dotted and
		    dash-double dotted lines respectively.}
	   \label{fig:bdQ}
	\end{figure}

	We can predict $Q_{nloc}$ from the power spectrum, assuming that the perturbations
	of the density field $\delta_m$ are small (i.e. the divergence of the velocity field $\theta_v$ 
	in equation (\ref{eq:g2}) is linear and therefore equal to $\delta_m$.
	 It also requires to assume that the legs of the considered triangles are large compared to the
	smoothing radius $R$ (large separation limit,
	i.e. in equation (\ref{eq:g2}), $W(Rq_{12})\simeq W(Rq_1)W(Rq_2)$). For such conditions
	the perturbation theory (hereafter also referred to as PT) offers the
	possibility to express the non-local three-point correlation $Q_{nloc}$ in terms of

        \begin{equation}
        \Gamma_{123} \equiv \left[\xi(r_{12}) + 3 \frac{\phi'(r_{12})}{r_{12}}\right] \left[\xi(r_{13}) + 3 \frac{\phi'(r_{13})}{r_{13}}\right] L_2(\cos\alpha_{12}),
        \label{eq:def_gamma}
        \end{equation}
        where 

        \begin{equation}
        \phi(r)=\int d^3 \mathbf{k} \frac{P(k)}{k^2}W^2(kR) \frac{\sin(kr)}{kr}
        \end{equation}
        and $\phi'(r)\equiv \frac{\dif \phi}{\dif r}(r)$. 
        One can show that 

        \begin{equation}
        Q_{nloc}(r_{12},r_{13},\alpha) = \frac{2}{3}\left\{ \frac{ \Gamma_{123}+ \Gamma_{312} + \Gamma_{231}}{\zeta^{mm}_H}  -1 \right\}. \\
        \label{eq:def_Qgamma}
        \end{equation}
        The angular dependence of the non-local component of the three-point halo auto- and
        cross-correlation functions is encoded in  equation (\ref{eq:def_gamma}) via the 
        second-order Legendre polynomial $L_2(\cos\alpha_{12})$. As shown by 
        \citet{bargaz} in their equation (8), at the tree-level and for large separations, the matter 
        three-point correlation can be expressed in the same way as expression (\ref{eq:def_Qgamma}).
        That is with respect to circular permutations of a function $\hat\Gamma_{123}$, expressed as 
        a monopole, a dipole and a quadrupole in $\cos(\alpha_{12})$ \citep[similar Legendre expansion
        have been used in Fourier space by][]{SB&S15}
        As a result, a non-local component, such as the tidal field $\mathcal{G}_2$ (equation (\ref{eq:g2})),
        modifies the amplitude of the quadrupole of $\hat\Gamma_{123}$ by an amount proportional to
        the non-local bias $g_2$ (see Appendix \ref{sec:nloc_bias}).
        
        Moreover, by comparing equations (\ref{eq:b1c2_q3auto}) and (\ref{eq:b1c2_q3cross}) one can see
        that quadratic and non-local contributions to the halo three-point correlation
        affect the cross-correlations by a factor $1/3$ less than the auto-correlation.
        The linear bias, in contrast, affects the auto- and cross-correlation equally. We will use this
        property to isolate the linear from the quadratic and non-local bias, as explained below.

        In the following we study non-local contributions to the halo bias in the  MICE-GC
        simulation. Ours measurements of the three-point correlation are based
        on the algorithm suggested in \citet{bargaz} \citep[see also][for studies of
        numerical effects in this algorithm and the impact of covariance between angular bins
         on the bias measurements]{paperI}. Errors are derived from $64$ cubical Jack-Knife samples.
        We first focus on the mass bin M2 at redshift $z=0.5$ to present our methods for
        extracting the parameters $b_1$, $c_2$ and $g_2$ from three-point correlations,
        which were computed using triangles with fixed legs of $r_{12}=32$ and $r_{13}=64 \ h^{-1}$Mpc.
        Afterwards we will present results for all mass samples and redshifts at various scales.
     
\subsubsection{Local bias}

        Our first method for measuring bias from three-point correlations is based on the local bias model ($g_2=0$).
        The linear and quadratic bias parameters are computed from the equations (\ref{eq:b1c2_q3auto})
        and (\ref{eq:b1c2_q3cross}) by fitting the $b_1$ and $c_2$ parameters which allows for mapping
        $Q_m$ into $Q_h$ and $Q_h^\times$, i.e. $Q_h=(Q_m+c_2)/b_1$ and
        $Q_h^\times=(Q_m+\frac{c_2}{3})/b_1$. The fitting procedure is explained
        in \citet{paperI}. The two estimations of the doublet $b_1$ and $c_2$ are respectively called
        $b_{Q}$, $c_{Q}$ and $b_{Q}{^\times}$, $c_{Q}^{\times}$. 
        In Fig. \ref{fig:q3cross_dmvshalo} we show how well a linear relation, expected from
        the local bias model, describes the mapping between the matter three-point correlation 
        and the three-point auto- and cross-correlation functions of haloes. However, we can see
        that the slope of the linear relation is different when considering auto- and cross-correlations,
        which indicates that the two methods deliver inconsistent results
        (see right panel of Fig. \ref{fig:q3cross_dmvshalo}) .
        As explained in \citet{chan12} this linear relation between matter and haloes might arise
        from a projection effect due to the fact that we neglect the non-local component $g_2 
        Q_{nloc}$.
        In Fig.~\ref{fig:q3cross_dmvshalo} we show that, if the contribution of $Q_{nloc}$
        is small compared to $Q_m$ (i.e. $g_2$ is small, see Fig. \ref{fig:b1-gamma2}),
        then they can indeed be approximately related by a linear relation.
        The ignored non-local contribution to halo three-point correlations
        can therefore be absorbed by $b_1$ and $c_2$, without substantially decreasing the
        goodness of the $Q_h$ and $Q^\times$ fits. The same effect
        has been shown in Fourier space by \citet{baldauf12} in their Fig. 1. This ignorance might lead to incorrect bias
        measurements, unless $g_2=0$.

\subsubsection{Non-local bias}

      	\begin{figure}
	   \centering
	   \includegraphics[width=75mm, angle = 270]{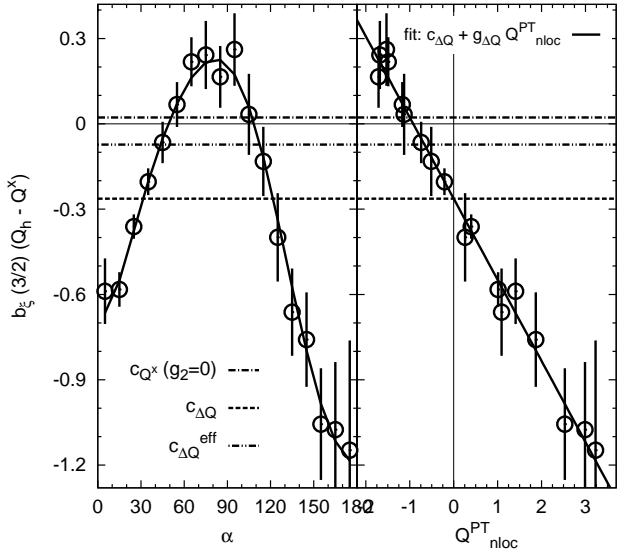}
	   \caption{
	         {\it Left:} The difference between the three-point auto- and cross-correlation
	         $Q_h$ and $Q^{\times}$ ($\Delta Q_{cg}$), multiplied with the 
	         linear bias from the two-point correlation (open circles) as a 
	         function of triangle opening angle $\alpha$ for M2 at $z=0.5$.
	         According to equation (\ref{eq:deltaqcg})
	         this expression is equivalent to $c_2+g_2Q_{nloc}$. If the bias function is quadratic and 
	         local, the measurements should correspond to the quadratic bias parameter $c_2$, as
	         measured from the halo-matter cross-correlation via equation (\ref{eq:b1c2_q3cross})
	         with $g_2=0$ (dash-dotted line). Fits from equation (\ref{eq:deltaqcg}), based on the PT
	         prediction for the non-local component, $Q_{nloc}$, are shown as thick solid line. 
	         The quadratic bias parameter, derived from this fit ($\Delta Q_{cg}$) is shown as 
	         dashed line. The corresponding effective quadratic bias parameter (equation (\ref{eq:c2eff}))
	         is shown as dashed-double dotted line. The three-point 
	         correlations were calculated from triangles with fixed leg of
	         $r_{12}=r_{13}/2 = 32 \ h^{-1}$Mpc.
	          {\it Right:} same as left panel, but showing the measurements versus the $Q_{nloc}$
	          prediction for each opening angle.}
	   \label{fig:dQnoloc}
	\end{figure}
	
	\begin{figure}
	   \centering
	   \includegraphics[width=75mm, angle = 270]{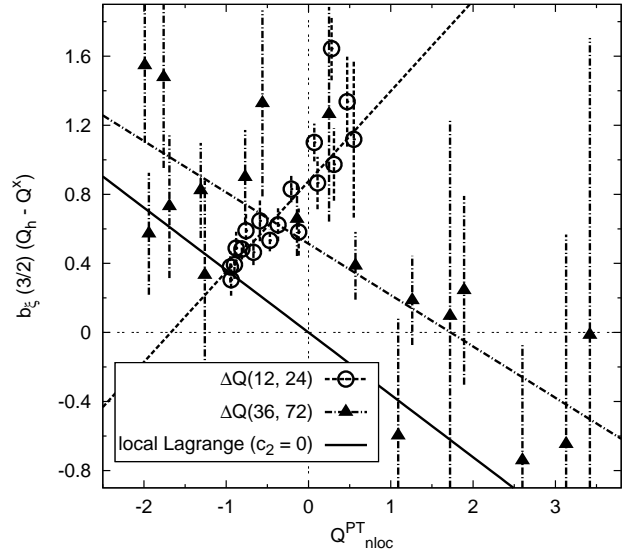}
	   \caption{
	         Same as right panel of Fig. \ref{fig:dQnoloc}, but for the halo mass sample M3 at
	         the redshift $z=0.0$. Results are shown for triangles with fixed legs
	         $r_{12}=r_{13}/2$ of $12$ and $36 \ h^{-1}$Mpc (open circles and filled triangles respectively).
	         Dashed and dashed-dotted lines are the corresponding fits to equation (\ref{eq:deltaqcg}),
	         used to derive the quadratic and non-local bias $c_2$ and $g_2$.
	         The black solid line corresponds to the local Lagrangian prediction for
	         $c_2+g_2Q_{nloc}$ with $c_2=0$ and $g_2=2 \gamma_2/b_1$, while
	         $\gamma_2 = -(2/7)(b_1-1)$. The $\gamma_2$ values from these fits are shown in
	         Fig. \ref{fig:b1-gamma2}.}
	   \label{fig:dQnoloc_scales}
	\end{figure}
        
       Our second method for measuring bias from three-point correlations is a new a approach,
       which combines auto- and cross-correlations. These two statistics can be 
       combined in two different ways which allow us to isolate the linear bias from
       quadratic and non-local contributions to the bias model.
       Both combinations take advantage of the fact that
        the linear bias, $b_1$, affects $Q_h$ and $Q_h^\times$ equally, which is
        not the case for the quadratic and non-local contributions, $c_2$ and $g_2$.
        The linear bias can be obtained by combining the equations (\ref{eq:b1c2_q3auto}) and 
        (\ref{eq:b1c2_q3cross}),
        
        \begin{equation}
        b_{\Delta Q} \equiv -2 \frac{Q_m}{\Delta Q},
        \label{eq:bdQ}
        \end{equation}
        where $\Delta Q\equiv Q_h-3Q_h^\times$. The interesting property of this linear bias estimator is 
        that it is independent from quadratic (local and non-local) contributions to the bias function. It can
        therefore be used to verify if such contributions are indeed the reason for deviations between
        linear bias estimations from two- and three-point correlations, as we speculated in \citet{paperI}. 
        Note that the relevant quantities involved in equation (\ref{eq:bdQ}) depend on the opening
        angle $\alpha$, so does the estimator $b_{\Delta Q}$. Hence, our final $b_{\Delta Q}$
        measurement is derived by fitting a constant to $b_{\Delta Q}(\alpha)$.
        The use of $\Delta Q$ has also the advantage that off-diagonal elements
        in the covariance matrix between different opening angles are smaller. This covariance is difficult
        to access and its Jack-Knife estimation can affect the bias estimation at the $5$\% level \citep{paperI}.

        We show $b_{\Delta Q}$ for M2 at $z=0.5$ in Fig. \ref{fig:bdQ} 
        at each angle probed by the three-point functions together with the corresponding 
        fit. In the same figure we also display the estimations for the linear bias,
        derived from three-point auto- and cross-correlations ($b_Q$ and 
        $b_{Q^\times}$, obtained from equations (\ref{eq:b1c2_q3auto}) and (\ref{eq:b1c2_q3cross}),
         assuming the local bias model, i.e. $g_2=0$).
        As a reference, we also include the linear bias measurements from the two-point cross-correlation,
        $b_{\xi}^\times$, which we consider to be a reliable estimate of the true linear bias
        (see Section \ref{sec:second_order_bias}). The comparison in Fig. \ref{fig:bdQ} reveals that the
        measurement and the fit of the hybrid bias estimator $b_{\Delta Q}$ are consistent with the
        reference $b_\xi$, while we see that the biases obtained from $b_{Q}$ and $b_{Q^\times}$
        are over estimating the linear bias. This result confirms our speculation in \citet{paperI}, that
        differences between $b_{\xi}$ and $b_{Q}$ are mainly due to a non-local term in the bias model.
        Furthermore we verified that also the magnitude of the overestimation is in agreement with
        what we expect from neglecting non-local contributions. However, as we 
        pointed out in this aforementioned study, linear bias measurements from three-point correlations
        can also be affected by numerical effects, originating from the algorithm employed for deriving
        $Q$ as well as shortcomings in the estimation of the $Q$ covariance between different angular bins.

        One can notice that, as expected in case of non-local bias,
        the overestimation is larger in case of the auto-estimator ($b_Q$) compared to the cross-estimator
        ($b_{Q^\times}$).  Note that \citet{Pollack12} found an opposite trend, analysing
        a different simulation in Fourier space with a different mass resolution and cosmology.
        Their linear bias measurements from the  bispectrum are closer
        to peak-background split predictions than measurements from three-point cross-correlations,
        while the predictions might be lower than the true linear bias \cite[see e.g.][]{paperIII}.        
        
	\begin{figure*}
	   \centering 
	   \includegraphics[width=150mm, angle = 270]{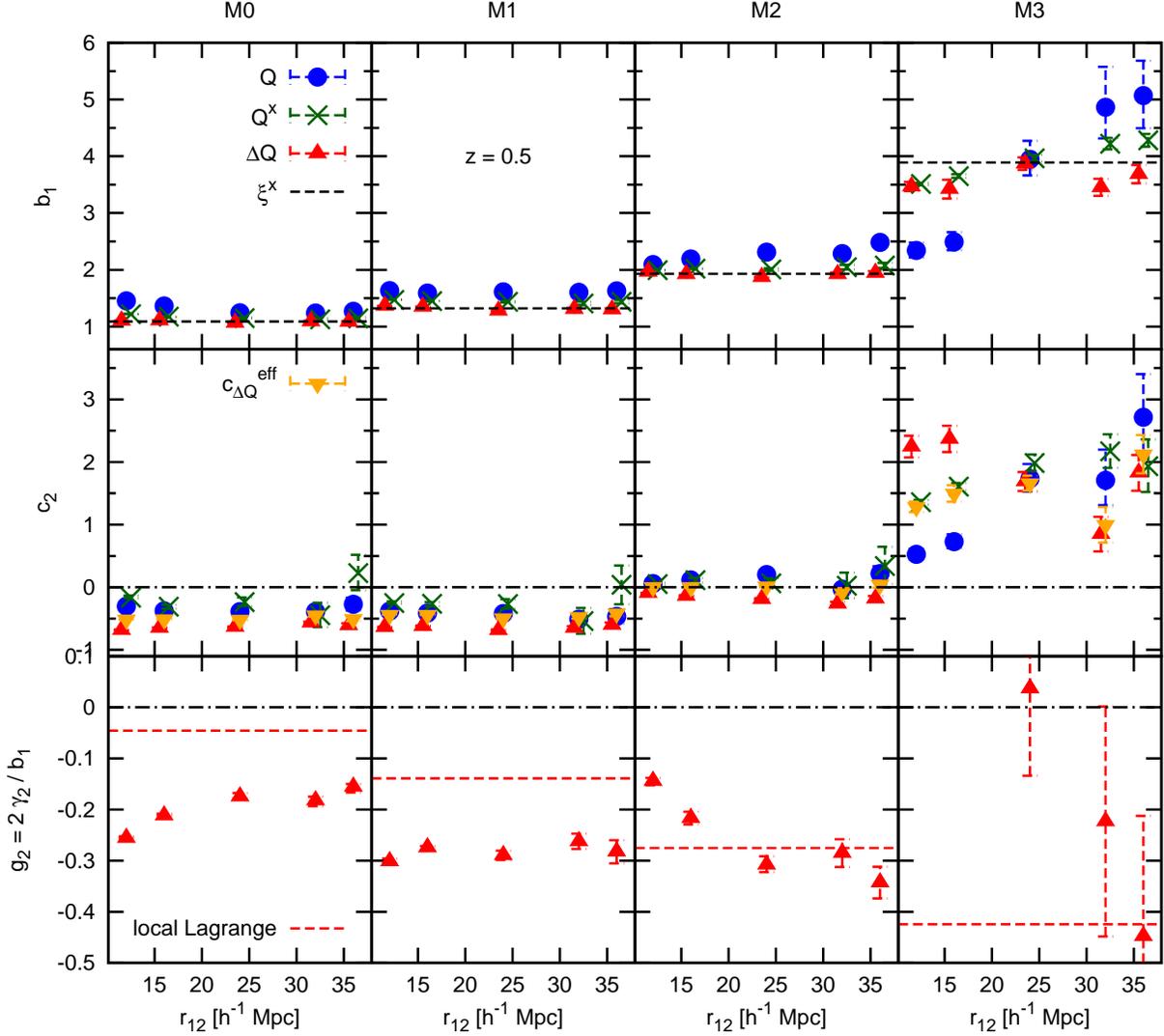}
	   \caption{Scale dependence of bias measurements from three-point 
	   correlations. Measurements were derived using triangles with $r_{13}/r_{12} = 2$ 
	   configuration, while scale of the smaller triangle leg, $r_{12}$, is denoted on the  x-axis
	   (slightly shifted for clarity).
	   Results are shown for the four mass samples M0-M3 (from left to right) at redshift $z=0.5$. The 
	   linear and quadratic bias (top and central panel respectively),
	   obtained from three-point auto- and cross-correlations using the local bias model 
	   ($g_2=0$) via the equations (\ref{eq:b1c2_q3auto}) and  (\ref{eq:b1c2_q3cross}) are 
	   shown as blue circles and green crosses respectively. The linear, quadratic and non-local bias,
	   measured using  combinations of three-point auto- and cross-correlations ($\Delta Q$, equation
	   (\ref{eq:bdQ} ) and (\ref{eq:deltaqcg})) are shown as red triangles. The 
	   linear bias is compared with reference measurements from the two-point
	   cross-correlation ($b_\xi^\times$), shown as black, dashed line. The 
	   quadratic bias measurements are compared with effective quadratic bias (equation 
	   (\ref{eq:c2eff})), shown as orange triangles. The non-local bias parameter 
	   $g_2 \equiv 2 \gamma_2 / b_1$ (bottom panel) is 
	   compared with predictions from the local Lagrangian model ($\gamma_2 = -(2/7)(b_1-1)$).
	   (red dashed line). Error bars denote $1\sigma$ uncertainties.}
	   \label{fig:b1c2gamma2}
	\end{figure*}

\subsubsection{Linear and quadratic terms}

        In order to further verify the presence of non-local contributions to the bias function
        we separate $Q_{nloc}$ from $Q_m$ by subtracting $Q_h^{\times}$ from $Q_h$ and define
        	
	\begin{equation}
		\Delta Q_{cg} \equiv Q_h-Q_{h}^\times=
		\frac{2}{3}\frac{1}{b_1}
		\left[c_{\Delta Q} + g_{\Delta Q} Q_{nloc}\right ],
		\label{eq:deltaqcg}	
	\end{equation}
 where $c_{\Delta Q}$ and $g_{\Delta Q}$ are the estimators of $c_2$ and $g_2$ in equations
  (\ref{eq:b1c2_q3auto}) and (\ref{eq:b1c2_q3cross}).
      If the bias function is quadratic in $\delta_m$ and local then the non-local bias parameter $g_2$ is 
      zero. Hence,
	
	\begin{equation}
              c_{\Delta Q} = b_1 \Delta Q_{cg} (3/2)
		\label{eq:cdeltaq}	
	\end{equation}
	should correspond to $c_2$, independently of the opening angle $\alpha$. In the left panel of
	Fig.~\ref{fig:dQnoloc},  we show $c_{\Delta Q}$, together with $c_2$, estimated from
	$Q_h$ and $Q^\times$  ($c_{Q}$ and $c_{Q^\times}$ respectively) from the local bias model 
	(equations (\ref{eq:b1c2_q3auto}) and (\ref{eq:b1c2_q3cross}), for $g_2=0$). In the same figure
	we also show $c_2$ derived from fitting ${\Delta Q_{cg}}$ taking into account a possible non-local
	bias $g_2$.
	The first important point is that the measured $c_{\Delta Q}$ shows a very clear angular 
	dependence, with a maximum at around $80$ degrees and a positive curvature.
	The local quadratic model therefore clearly fails in describing the impact of
	bias on three-point correlations.

      The right panel of Fig.~\ref{fig:dQnoloc} displays the measurements of $\Delta Q_{cg}$ with
      respect to the tidal, non-local component of the three-point function $Q_{nloc}$, as predicted by
      equation (\ref{eq:def_Qgamma}). It shows that measurements are compatible with a linear relation
      between $\Delta Q_{cg}$ and $Q_{nloc}$, which is expected in the second order non-local bias model.
      It therefore confirms the presence of non-local bias due to
      tidal forces ($\mathcal{G}_2$) in the matter field.  
     For such a linear relation, the value of $\Delta Q_{cg}$ corresponds to the second-order bias
      $c_{\Delta Q}$ when $Q_{nloc} = 0$, while its slope provides a direct insight to the non-local bias $g_2$.
      Finally, by comparing the various estimates of the second-order local bias $c_2$, one can see 
      that, when assuming a local bias, the results differ significantly from the ones obtained when
      taking into account a possible non-local component into the biasing relation.
      However, the estimation of $c_{\Delta_Q}$ and $g_2$ might be affected by shortcomings
      of the PT predictions for $Q_{nloc}$.
      
	  To verify how well the PT predictions describe the measurements at different scales  we show the
	  relation between $\Delta Q_{cg}$ and $Q_{nloc}$ derived from triangles with fixed legs $r_{12}=r_{13}/2$
	  of $12$ and $36 \ h^{-1}$ Mpc respectively in Fig. \ref{fig:dQnoloc_scales} for the mass samples M3 at $z=0.0$,
	  for which we find a large non-local bias amplitude. At large scales ($r_{12}=36 \ h^{-1}$ Mpc) the slope of the
	  measured $\Delta Q_{cg}$ - $Q_{nloc}$ relation is comparable with the local Lagrangian prediction. Interestingly
	  at small scales the $\Delta Q_{cg}$ - $Q_{nloc}$ relation is also linear, while the slope has the opposite sign
	  than at large scales. The linearity at small scales indicates that higher-order terms enter the $Q$ in a similar
	  way as second-order non-local contributions to the bias function. This suggests that linear bias measurements
	  can be improved by using the prediction for $Q_{nloc}$, while using the local Lagrangian prediction for the
	  non-local bias $g_2$ is only appropriate at extremely large scales.
	
	\subsubsection{Scale dependence}

      Based on the methods for measuring linear, second-order and non-local bias 
      from $Q$ and $Q^\times$ ($b_{Q}$, $b_{Q^\times}$,
      $b_{\Delta Q}$, $c_{Q}$, $c_{Q^\times}$ and $c_{\Delta Q}$),
      which were presented above, we now apply our analysis 
      to each of the mass samples M0-M3 at $z=0.5$. We study the scale dependence of our 
      results as before by varying the size of the triangle leg $r_{12}$ between
      $12 \ h^{-1}$Mpc and $36 \ h^{-1}$Mpc) while fixing $r_{12} / r_{13} = 1/2$.
      The various bias estimations are presented in Fig. \ref{fig:b1c2gamma2} for 
      different triangles sizes, defined by the length of $r_{12}$.
      
      From the comparison between the different bias estimations we draw similar conclusions
      as from the example of M2 at $z=0.5$. On linear scales (sufficiently large triangles) the linear
      bias parameters obtained from each method reaches a regime in which they become scale
      independent. However, they do not converge to the same value. In case of the linear bias,
      only $b_{\Delta Q}$ is in agreement with $b_{\xi}^\times$, while $b_{Q}$ and $b_{Q^\times}$
      overestimate the linear bias; this overestimation is stronger in the case of $Q$ than $Q^\times$.

      The scale dependence shown for the high mass bin M3 in Fig.~\ref{fig:b1c2gamma2}
      shows that, if the analysis is performed at too small scales, then one can measure a positive
      non-local bias, while it is in reality negative (as we see from the results derived at large scales).
      We indeed verified that for highly biased tracers and small triangles, the curvature of $c_{\Delta Q}$
      flips from positive to negative. This scale dependence indicates a domination of the signal by
      non-linear terms in these cases. However, for lower mass samples and large scales, for which we
      expect non-linear contributions to converge to zero, we still see a strong angular dependence
      of $c_{\Delta Q}$, which speaks for the presence of non-local bias contributions. Hence, we have
      shown that all the mass sample used in the present analysis exhibit a detectable non-local component.
     
     In case of estimators of the second-order bias $c_2$ (middle panel of 
     Fig.~\ref{fig:b1c2gamma2}) we observe the same tendency, however we do not have a reference
     estimate. As a result we shall be more confident in the estimation of $c_2$ coming from
     $c_{\Delta Q}$.  We compare the latter to measurements from third-order moments
     in Section \ref{sec:results}. A comparison with $c_2$ derived from various methods
     will be presented in Bel et al. (in preparation).
     
     The bottom panel shows that each mass sample comprises non-local bias
     which significantly differs from the local model $\gamma_2=0$. These 
     measurements therefore constitute the first detection of non-local bias in 
      configuration space. In the case of M0 and M1 the amplitude of the
     non-local bias strongly differs from the local Lagrangian biasing relation
     between the halo and matter field \citep{mo96}.
	
\subsubsection{Non-local to linear bias relation}	
	
     \begin{figure}
	   \centering
	   \includegraphics[width=110mm, angle = 270]{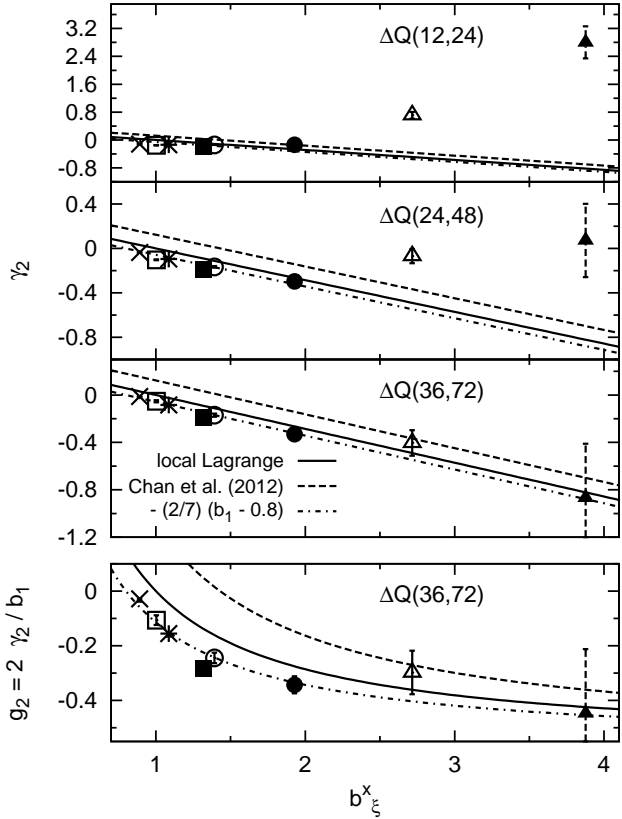}
	   \caption{Non-local bias parameters $\gamma_2$ and $g_2=2\gamma_2/b_1$ versus
	   the linear bias from the two-point cross-correlation, $b^\times_\xi$. 
	   Measurements, derived from $\Delta Q_{cg}=(Q_h-Q^\times)$ via equation (\ref{eq:deltaqcg}) are
	   shown as symbols with $1\sigma$ errors. Results are shown for $Q$ measurements 
	   based on triangles with fixed legs of $r_{12}=12$, $24$ and $36 \ h^{-1}$Mpc
	   and $r_{12} = r_{13}/2$ configurations (from the top to the bottom).
	    Crosses, open squares, open circles and open triangles
	   show the non-local bias, measured for the mass 
	   samples M0, M1, M2 and M3 respectively at redshift
	   $z=0.0$. Stars, closed squares, closed circles and closed triangles show the corresponding M0-M3 
	   measurements at $z=0.5$.
	   The measurements for $r_{12}=36 \ h^{-1}$Mpc are approximated with
	   $\gamma_2 = -(2/7)(b_1-0.8)$ (black dash-dotted lines) and compared with
	   the local Lagrangian prediction ($\gamma_2 = -(2/7)(b_1-1)$) as well as to a fit 
	   given by \citet[][$\gamma_2 = -(2/7)(b_1-1.43)$]{chan12}, shown as solid and dashed black lines respectively.}
	   \label{fig:b1-gamma2}
      \end{figure}	

      Following \citet{chan12}, we compare our non-local bias measurements to the linear bias derived
      from the two-point cross-correlation in the top panels of Fig.~\ref{fig:b1-gamma2}.
      This comparison includes measurements at redshift $0.0$ and $0.5$ which are based on triangles
      with $r_{12}/r_{13}=1/2$ configurations.
      For very large triangles ($r_{12}=36 \ h^{-1}$Mpc) our results indicate a linear relation between
      the non-local $\gamma_2$ and the linear $b_1$ bias, as expected for the local Lagrangian biasing.
      However, the amplitude of this relation lies below the local Lagrangian prediction, which is the
      opposite of what was reported by \citet{chan12}.
      Some work is currently ongoing, aiming to explain whether these
      differences result from the fact that \citet{chan12} conduct their measurements 
      using the Bispectrum in Fourier space, while we employ the reduced 
      three-point correlation in configuration space. A further contribution to the 
      discrepancies could arise from differences in the simulation, such as mass 
      resolution effects, or differences between cosmological
      parameters. 

      As a matter of fact, our measured $b_1-\gamma_2$ relation shows the same tendency as
      those of \citet{SC&S13}, \citet{baldauf12} and \citet{Saito14} who also find (in Fourier space) $\gamma_2$
      to be below the
      local Lagrangian prediction. The simulations employed in the two latter studies are based on
      cosmologies similar to MICE-GC, while \citet{chan12} study a simulation with a initial power
      spectrum which is significantly different in terms of its normalisation $\sigma_8(z=0.0)$
      and spectral index $n_s$.
      Note also that the departures from the local Lagrangian prediction in Fig.~\ref{fig:b1c2gamma2} are 
      strongly scale dependent for highly biased samples ($b_{\xi} \gtrsim 2$), which
      indicates the presence of non-linear contamination to $Q_m$ and $Q_{nloc}$ \citep[e.g.][]{Saito14}.

	\subsection{Bias from third-order moments $b_{\tau}^{\times}$}\label{sec:btau}
	
	An alternative approach for exploring the higher-order statistical properties of 
	the large-scale structure are one- and two-point third-order statistics,
	respectively referred to as skewness and reduced correlator.
	They can be understood as the one- and two-point limits of
	the reduced three-point correlation \citep[e.g.][]{gaztafrie94}.
	Since these statistics are isotropic, they do not provide access to the full third-order hierarchy
	probed by the three-point correlation (i.e. the shape of the large-scale structure). However,
	this section will show that this apparent disadvantage leads to a cancellation of non-local bias
	which is useful for linear bias measurements, as we discuss later.

     The auto-skewness and the reduced auto-correlator of the matter density fluctuations
     $\delta_m$ are respectively defined as
      
      \begin{equation}
      S_3 \equiv\frac{\langle \delta_m^3\rangle}{\sigma^4},
     \label{eq:S3auto}
      \end{equation}
      and
      
      \begin{equation}
      C_{12}(r)\equiv\frac{\langle\delta_{m,1}\delta_{m,2}^2\rangle}{ \sigma^2 \xi(r) }
      \label{eq:C12auto}
      \end{equation}
      \citep[see][]{g86, bernardeau96}. As in the previous subsection $\sigma$ 
      and $\xi$ refer to the matter field. The auto-skewness and the reduced auto-correlator
      for halo density fluctuations, $\delta_h$, are defined analogously to equation
      (\ref{eq:S3auto}) and (\ref{eq:C12auto}).
      The skewness is directly related to the asymmetry of the one-point
      probability distribution of $\delta_m$, while the correlator tells how the quadratic field is correlated
      with itself on a given scale $r$. Their properties have been extensively investigated in literature
      \citep[e.g.][]{bernardeau96,GFC,bm}.
      %
       As for the two-point and three-point correlation we also study the cross-skewness
       $S_{3,h}^\times$ and the cross-correlator $C_{12,h}^\times$ in order to investigate
       the impact of non-linearities, non-local bias as well as shot-noise on the measurements.
       We define these quantities as
	
	\begin{equation}
      S_{3,h}^\times  \equiv  \frac{\langle\delta_h\delta_m^2\rangle}{\sigma_{\times}^4}
     \label{eq:S3cross}
      \end{equation}
      and
      
      \begin{equation}
      C_{12,h}^\times  \equiv \frac{\langle\delta_{h,1}\delta_{h,2}\delta_{m,2}\rangle}{\sigma_{\times}^2\xi_h}.
	 \label{eq:C12cross}
      \end{equation}
      As in previous sub-sections the indices $h$ and $m$ refer, respectively, to the halo and 
      matter density contrast.
	\begin{figure*}
	\includegraphics[width=80mm,angle=0]{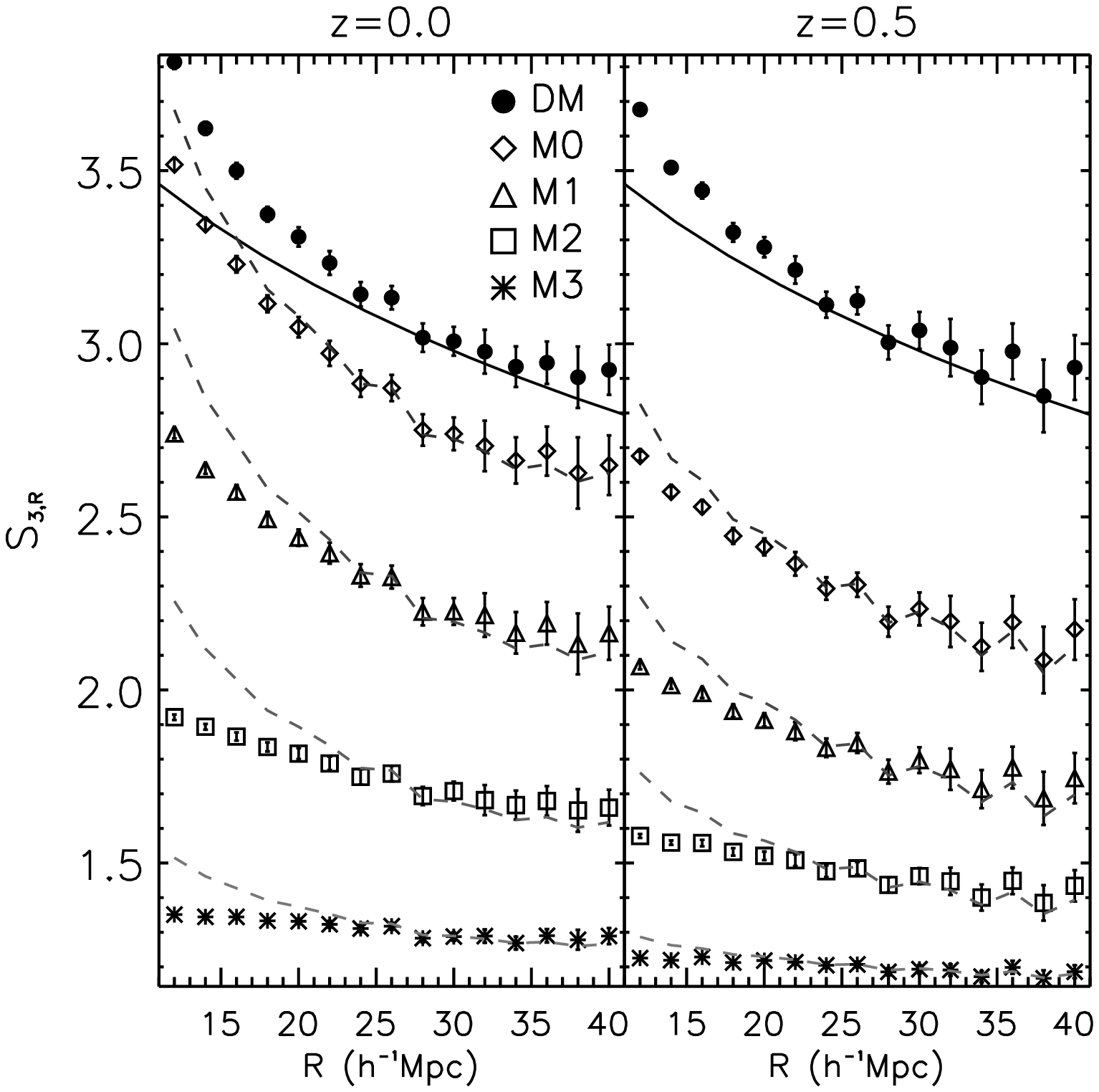}
	\includegraphics[width=80mm,angle=0]{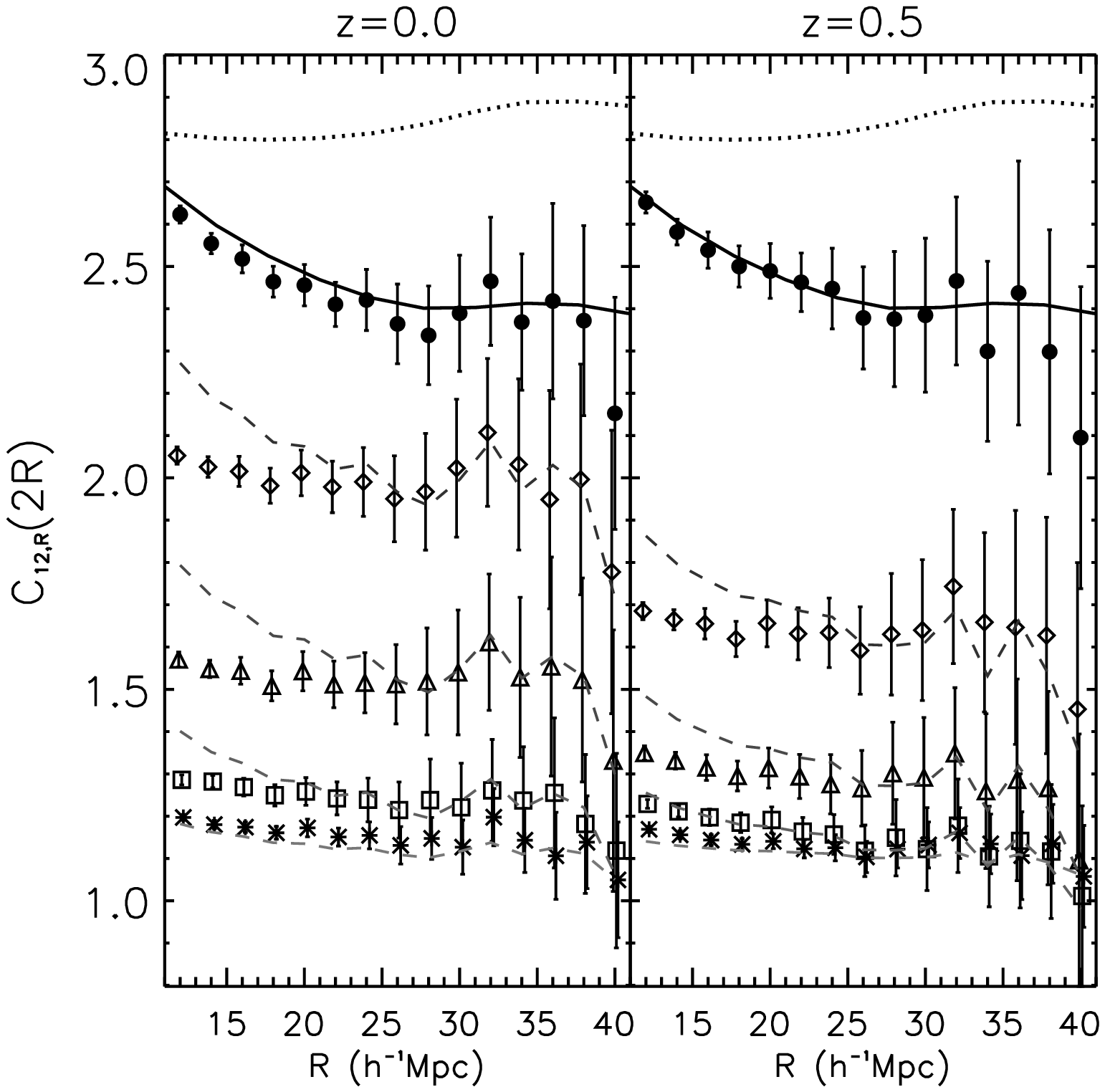}
	\caption{\small  {\it Left:} reduced cross Skewness $S_3^\times$, measured at the two
	redshifts $z=0.0$ and $z=0.5$ for matter (filled circles), and the four halo mass samples M0-M3
	(diamonds, triangles, squares and stars respectively). {\it Right:}
	corresponding measurements for the reduced cross-correlator $C_{12}$.  On each panel we
	display the tree-level PT predictions for $S_3$, equation (27)
        in \citet{paperI}. For $C_{12}$ we show two predictions. The
        dotted line displays the one from equation (28)  in \citet{paperI}
        and the solid black line shows the one taking into account the
        small separation contribution (Bel in preparation).
        We also display in grey dashed
	lines the fits to $S_{3}^\times$ and $C_{12}^\times$ from the equations (\ref{eq:S3crossS3m}) and
	(\ref{eq:C12crossC12m}).}
	\label{fig:S3C12}
	\end{figure*}
       
       For measuring the auto skewness $S_3$ and the auto correlator $C_{12}$ we follow \citet{bm} 
       by setting up a regular grid of spherical cells of radius $R$ and counting the number of 
       objects (haloes or dark matter particles) per cell. After assigning a number density contrast $\delta_N$ to each grid cell we
       derive the auto-skewness $S_3$ as
      
      \begin{equation}
	S_{3,N}=\frac{\langle\delta_N^3\rangle-3\langle\delta_N^2\rangle\bar N^{-1}+2\bar N^{-2}}{ (\langle\delta_N^2\rangle-\bar N^{-1})^2}.
	 \label{eq:S3autoestim}
      \end{equation}
      In order to estimate the reduced correlator $C_{12}$, we consider the density contrast
      $\delta_{N,1}\equiv\delta_N$ for each grid cell. We then place an isotropic
      distribution of cells at distance $r$ around the central cell and
      assign the number density contrast $\delta_{N,2}$ to each of these surrounding cells.
      %
      By averaging over all grid points in the simulation volume we estimate the correlator as

      \begin{equation}
	C_{12,N}
	=\frac{ \langle\delta_{N,1}\delta_{N,2}^2\rangle-2\langle\delta_{N,1}\delta_{N,2}\rangle\bar N^{-1} }{  \langle\delta_{N,1}\delta_{N,2}\rangle(\langle\delta_N^2\rangle-\bar N^{-1}) },
	 \label{eq:S3autoestim}
      \end{equation}
      where $\bar N$ is the average number of halo or dark-matter particles per cell
      in the simulation box.
      These estimators are corrected for shot-noise, assuming the local Poisson process approximation
      \citep{Layser}. Note that, in order to be able to handle the large number of dark matter particles,
      we use only $1/700$ of the total number of particles in the dark matter simulation output.
      In principle this neglection of particles introduces additional shot-noise errors but we have
      tested that it does not affect the measurements.
  
     Following the same method,  the cross-skewness $S_{3,h}^\times$ is estimated as

      \begin{equation}
      S_{3,h}^\times=\frac{ \langle\delta_{N,h}\delta_{N}^2\rangle - \langle\delta_{N,h}\delta_N\rangle\bar N^{-1} }{ (\langle\delta_{N,h}\delta_N\rangle^2 },
      \label{eq:S3crossestim}
      \end{equation}
      where the sub-index $h$ again refers to halo density field.  Note that we correct the cross-skewness
      differently for shot-noise than the auto-skewness, since the former depends on the matter-field
      at the power of $2$. Finally, since the cross-correlator does not involve any powers of the same
      density field larger than unity, the reduced cross-correlator $C_{12,h}^\times$
      is expected to be
      insensitive to Poisson shot-noise.  Hence we estimate it directly from equation (\ref{eq:C12cross}).
       The corresponding errors are estimated with Jack-Knife sampling, using $64$ cubical cells.

      Our measurements of the skewness and the reduced correlator are presented 
      as a function of smoothing scale $R$ in Fig. \ref{fig:S3C12} for the redshifts $z=0.0$ and $z=0.5$. 
      As in Section \ref{sec:second_order_bias}  the scale dependence is 
      explored by varying the smoothing scale $R$ while we fix the ratio $r/R=2$ in case of the
      correlator. In the same figure we also show the PT predictions for the dark matter field.
      %
       The skewness of the dark matter field appears to be redshift independent in the linear regime
       ($R\gtrsim20h^{-1}$Mpc), which is in good agreement with the tree-level PT prediction \citep{bernardeau92}.
       At smaller smoothing scales the redshift dependence becomes more significant which has been
       shown by \citet{F&G98} to be due to non-linear loop corrections. The reduced correlator of the
       dark matter field shows only a weak redshift dependence, even for small smoothing radii
       ($R \lesssim 10h^{-1}$Mpc). In \citet{paperI} we have shown that the measured matter correlator
       is in disagreement with the tree-level PT prediction (dotted blue line in the right panel of
       Fig. \ref{fig:S3C12}). We argued that this is due to the fact that the standard tree-level prediction
       \citep{bernardeau96, bm} neglects the small separation contributions. We are indeed looking at
       the correlation between adjacent spheres. In the right panel of  Fig.~\ref{fig:S3C12} we display 
       also the tree-level prediction (magenta dotted line) which takes into account the small separation
       contribution (Bel in preparation) arising from the mode-coupling induced by the smoothing process.
       The observed agreement confirms that for small separations it is mandatory to take into account 
       the smoothing in a proper way.

       In Fig. \ref{fig:S3C12} we also show for the first time measurements of the
       halo-matter cross-skewness and  cross-correlator. Comparing these measurements to our 
       results for the auto-skewness and the auto-correlator, presented in \citet{paperI},
       we notice two qualitative differences. The first difference is that the 
       cross-skewness for the highest mass sample M3 increases at small scales,
       while we found an opposite trend only for M3 in case of the 
       auto-skewness. The latter measurement should be more affected by shot-noise
       than the corresponding cross-skewness one or than measurements at 
       lower mass samples; this is due to the low number density of M3 haloes. Furthermore the impact
       of shot-noise increases at small scales, where the cell volumes are small as well.
       The decrease of the M3 auto-skewness at small scales might therefore be attributed to
       a failure of Poisson shot-noise correction, indicating the presence of non-Poisson shot noise 
       for highly biased samples.
       
       The second difference is that the amplitudes of the cross-skewness and the cross-correlator
       increase monotonically with the halo mass. We did not find such a monotonic behaviour
       for the auto-skewness and the auto-correlator in \citet{paperI} which has been also confirmed by
       \citet{AB&L08}. This difference can be explained by the fact that the bias
       parameters affect the auto- and cross-statistics differently, as we demonstrate in 
       the following.
           
\subsubsection{Bias relations}

       The skewness and the correlator of the matter field can be related to the corresponding halo and
       halo-matter cross-statistics via the non-local bias model. In analogy to the three-point correlation,
       one can express, at the leading order, the cross-statistics between haloes and matter as function of the
       bias parameters by inserting the bias functions from equation (\ref{eq:def_biasmodel}) into 
       the definitions  (\ref{eq:S3auto}) - (\ref{eq:C12cross}). In order to
       take into account the non-local component, we further assume that perturbations are well
       described by tree-level PT. 
       It leads to the following expressions for the skewness, correlator, cross-skewness and
       cross-correlator of haloes

	\begin{eqnarray}
       S_{3,h} & \simeq & \left[ S_3 + 3(c_2 - \frac{2}{3}g_2) \right]/b_1   \label{eq:S3hS3m} \\
	C_{12,h} & \simeq & \left[ C_{12} + 2( c_2 - \frac{2}{3}g_2) \right] /b_1  \label{eq:C12hC12m} \\
	S_{3,h}^\times & \simeq & \left[ S_3 + c_2 - \frac{2}{3}g_2 \right] /b_1   \label{eq:S3crossS3m} \\
	C_{12,h}^\times & \simeq & \left[ C_{12} + c_2 - \frac{2}{3}g_2 \right] /b_1  \label{eq:C12crossC12m}
	\end{eqnarray}
      (see Appendix \ref{sec:nloc_bias}). 
      On can notice the similarity with the corresponding equations for the three-point 
      auto- and cross-correlations (\ref{eq:b1c2_q3auto}) and  (\ref{eq:b1c2_q3cross}).
       The contribution of the second-order biases (local or non-local) to the cross-skewness and
       cross-correlators is three times smaller than in case of the auto-skewness and two time smaller
       than in the auto-correlator.
       
       In contrasts to the three-point correlations the second-order non-local contributions,
       described by equation (\ref{eq:g2}), can be taken into account as an effective second-order local 
       bias,
      \begin{equation}
      c_2^{\mathrm{eff}}=c_2 - \frac{2}{3}g_2,
      \label{eq:c2eff}
      \end{equation}
      which consists of a local ($c_2$) and a non-local ($g_2$) contribution. This absorption of
      non-local contributions by $c_2^{\mathrm{eff}}$ results from the isotropy of spherically
      averaged quantities, such as the skewness or the correlator \citep[][already presented this
      effective local description of non-local biasing for spherically symmetric matter
      perturbations]{chan12}.
     Hence, we do not expect the estimation of the linear bias via the equations 
      (\ref{eq:S3hS3m})-(\ref{eq:C12crossC12m}) to be significantly affected by non-local
      contributions to the bias model. However, such non-local contributions will modify
      systematically the estimation of the quadratic bias parameter $c_2$. We will discuss the impact on
      the bias estimations in Section \ref{sec:results}.

       By varying the bias parameters in equation (\ref{eq:S3crossS3m}) and (\ref{eq:C12crossC12m}) 
       we can fit the measurements of the auto-skewness and auto-correlator for matter to the
       measured halo-matter cross-skewness and cross-correlator. We see in Fig. \ref{fig:S3C12} that
       these fits are in reasonable agreement with the measurements for $R\gtrsim 20 \ h^{-1}$Mpc.
       At smaller scales the fits lie typically above the measurements. This latter discrepancy is lower
       than in the case of the auto-statistics, presented in \citet{paperI}. We therefore conclude that
       non-linearities  and non-local contributions have less impact on the third-order
       cross-statistics than on the corresponding auto-statistics.
      
      In order to measure the linear and the quadratic bias parameters we combine equations
      (\ref{eq:S3crossS3m}) and (\ref{eq:C12crossC12m}) in a way which allows us to 
      measure linear and quadratic contributions (local or non-local) to the bias model
      independently from each other
      
      \begin{eqnarray}
	b_{\tau}^{\times}\!\!\! & \equiv & \!\!\frac{{\displaystyle \,S_{{3}}-\,C_{{12}} }}{{\displaystyle \,S_{{3,h}}^\times -\,C_{{12,h}}^\times }}\equiv \frac{\tau^\times}{\tau_h^\times} \label{eq:b1taucross} \\
	c_{\tau}^{\times}\!\!\!& \equiv & \!\!\frac{{\displaystyle S_{{3}}C_{{12,h}}^\times- C_{{12}}S_{{3,h}}^\times }}{{\displaystyle  \tau_h^\times }}  \label{eq:c2taucross}.
	\end{eqnarray}
      This type of $S_3$ and $C_{12}$ combinations has been suggested for the 
      auto-skewness and the auto-correlators by \citet{Szapudi98} and \citet{bm} 
      and was further studied by \citet{paperI}.
      
\subsubsection{Bias measurements}

\begin{figure*}
      \centering
       \includegraphics[width=220mm,angle=0]{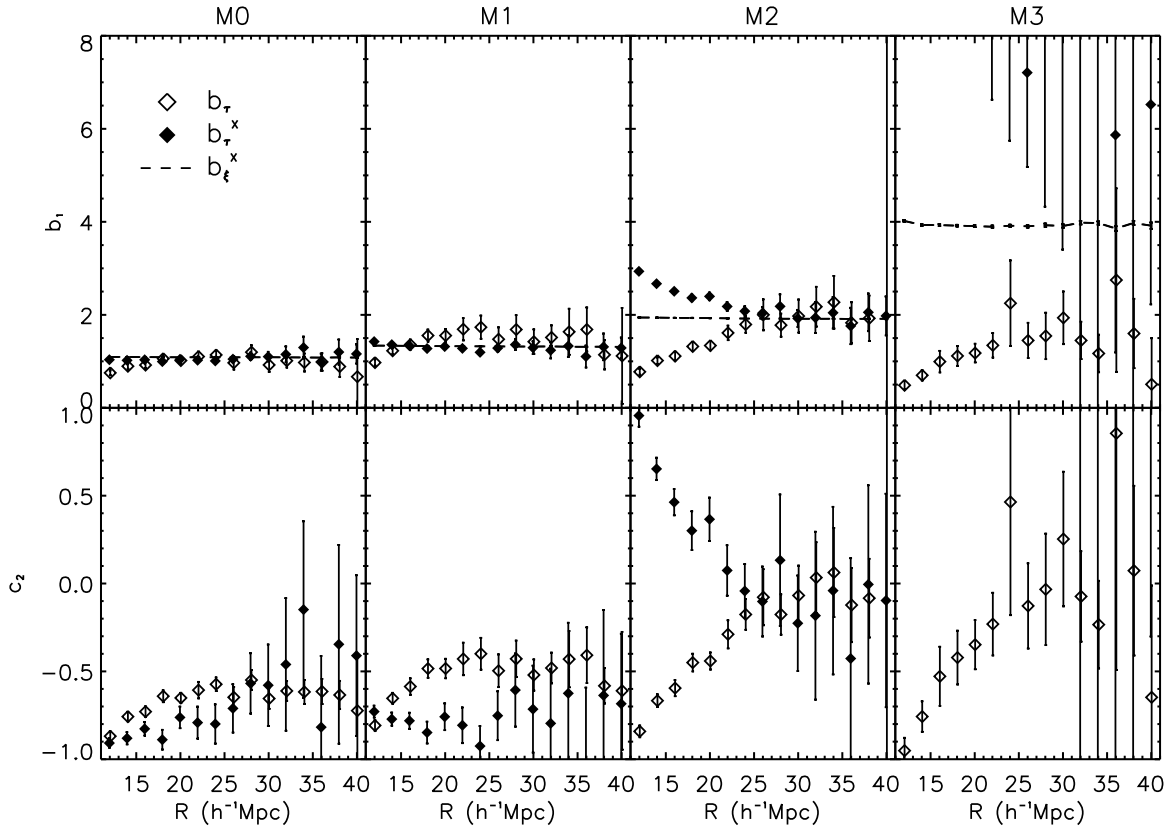}
	\caption{ {\it Top:} Linear bias obtained from the
            $b_\tau$ and $b_\tau^\times$ estimators (respectively open and filled diamonds) compared to
            the reference linear bias from the two-point
            cross-correlation (dashed line) with respect to the smoothing
            scales $R$ and at $z=0.5$ for the mass samples  M0-M3 (Table \ref{table:halo_masses}).
            {\it Bottom:} Quadratic reduced bias parameter $c_2$ estimated with $c_\tau$ estimator
            (diamonds) at $z=0.5$. Error bars denote $1-\sigma$ uncertainties.
            }
	\label{fig:b1c2mznl}
	\end{figure*}

      Our measurements for the bias parameters, obtained from equations (\ref{eq:b1taucross}) and
	(\ref{eq:c2taucross}), are displayed in Fig. \ref{fig:b1c2mznl} with respect to the smoothing
	radius $R$. 
	As for the measurements of $C_{12}$ we set the distance between two spherical grid cells
	with radius $R$ to $r=2R$.
	The measurements were performed for the four mass bins M0-M3
	at redshift $0.0$ and $0.5$. We compare our results with corresponding 
	measurements from the auto-skewness and the auto-correlator, $b_\tau$ and $c_\tau$, presented in \citet{paperI}.
	%
	The upper panel shows the linear bias measurements together with the
	reference measurements from the two-point cross-correlation
	(equation (\ref{eq:bxicross})), which we consider to be a robust estimate of 
	the true linear bias (see Section \ref{sec:second_order_bias}).
      We find that, for  all masses, $b_\tau^\times$ is less sensitive
      than $b_\tau$ to the considered
      scale.
      This finding is consistent with our previous conclusion from Fig.~\ref{fig:S3C12}
      that cross-statistics must be less affected by non linearities.
      The scale dependence is stronger for higher masses which can be attributed to 
      non-linear and non-local terms in the bias functions \citep{Saito14}. This interpretation is
      supported by the fact that we find a better agreement between  $b_\xi$ and $b_\tau$ at redshift
	$0.5$ than at redshift $0.0$.

      Our second-order bias estimations $c_\tau$ and $c_\tau^\times$ are shown 
      in the bottom panel of Fig. \ref{fig:b1c2mznl}.
      It is worth noticing that, as opposed to linear bias estimators,
      the $c_\tau^\times$ estimator is affected by a larger scatter in the measurement than $c_\tau$.   
      This can be explained by lower sensitivity of the cross-statistics
      ($S^\times_{3,h}$ and $C^\times_{12,h}$) to non-linear and non-local contributions
      to the bias functions (i.e. $c_2^{\mathrm{eff}}$), compared the auto-statistics
       ($S_{3,h}$ and $C_{12,h}$). This is indeed related to their monotonic behaviour with
       respect to the considered mass sample (see Fig.~\ref{fig:S3C12}); the main dependence 
       comes from the linear bias. The combination of $S^\times_{3,h}$ and $C^\times_{12,h}$
      to obtain $c_\tau^\times$ therefore provides weak and noisy constrains on $c_2^{\mathrm{eff}}$.
      
      
      The fact that for highly biased tracers (M3) any non-local component in the bias relation 
      is unable to explain why the two estimators $b_\tau$ and $b_\tau^\times$ do not
      converge to the $b_\xi^\times$ estimator indicates the presence of a velocity bias, which has been
      shown to appear for very massive haloes \citep[see][]{chan12}. This kind of contribution
      would add a dipole to the biasing relation which  could modify significantly the $b_\tau$ and $b_\tau^\times$ estimators.

\section{Results}\label{sec:results}
          
          In the previous section we studied measurements of the linear, quadratic and
          non-local bias parameters ($b_1$, $c_2$ and $g_2$ respectively). These measurements
          were derived from three different methods which are all based on third-order statistics
          of halo- and matter density fluctuations.
          
          Two of these methods employ three-point auto-and cross-correlations.
          The first method is to compare the three-point cross-correlation with the
          three-point matter auto-correlation to derive the linear and quadratic
          bias parameters via equation (\ref{eq:b1c2_q3cross}). This approach is based
          on the assumption of a local bias model. The linear and quadratic bias parameters, derived from
          this method are called $b_Q^\times$ and $c_Q^\times$ respectively.
          The second method is to use particular combinations of three-point auto- and 
          cross-correlations. The linear bias parameters, derived this way via equation
          (\ref{eq:bdQ}), $b_{\Delta Q}$, are independent of any quadratic contributions
          (local or non-local) to the bias function, as explained in the previous section.
          The quadratic and non-local bias parameters
          are obtained simultaneously by fitting predictions for the non-local component
          of the three-point correlation function, $Q^{PT}_{nloc}$, to $\Delta Q_{cg}$,
          defined in equation (\ref{eq:deltaqcg}). The quadratic parameters from such
          measurements is called $c_{\Delta Q}$.
          
          The third method is to use a combination of the halo-matter cross-correlator and cross-skewness
          with the corresponding auto-statistics for the matter field. The linear bias, derived
          from this so-called $\tau^\times$ estimator via equation (\ref{eq:b1taucross}) is called 
          $b^\times_\tau$. The effective quadratic bias parameter, obtained from equation (\ref{eq:c2taucross})
          is called $c^\times_\tau$. We demonstrated in the previous section that this 
          effective quadratic bias ($c_2^{\mathrm{eff}}$) consists of the quadratic bias and a non-local 
          contribution, which cannot be distinguished from each other with the $\tau$ method
          (see equation \ref{eq:c2eff}). This feature is a consequence of the isotropy 
          of the skewness and the correlator. Since we have shown that cross-statistics are weakly
          sensitive to the effective second order bias and therefore more 
          sensitive to shot-noise we will use instead the $c_\tau$ estimator. 
         For clarity, in Table \ref{table:definitions} we summarize our
           estimators and the notations that we use.

      \begin{table}
      \centering
        \caption{Bias estimators, definitions and notations}
        \label{table:definitions}
        \begin{tabular}{c  c c }
       Symbol  & Definition  &  Equation \\
           \hline 
             $b_\xi^\times$	&	$ \xi^\times/\xi $	&	Eq.(\ref{eq:bxicross})   \\
             $b_Q$	&	$Q_h =\frac{1}{b_Q}\bigl\{ Q_m+ c_Q \bigr\}$	&	Eq.(\ref{eq:b1c2_q3auto})	  \\
             $b_Q^\times$	&	$Q_{h}^\times = \frac{1}{b_Q}\bigl\{ Q_m+ \frac{c_Q}{3} \bigr\}$	&	Eq.(\ref{eq:b1c2_q3cross})   \\
             $b_{\Delta Q}$	&	$b_{\Delta Q} \equiv -2 \frac{Q_m}{ Q_h-3Q_h^\times}$	      &	Eq.(\ref{eq:bdQ})	    \\
             $b_{\tau}$	&	$\frac{\tau}{\tau_h}$	      &	\citet{paperI}	    \\
             $b_{\tau}^\times$	&	$\frac{\tau^\times}{\tau_h^\times}$	      &	Eq.(\ref{eq:b1taucross})	    \\
             $c_{Q}$	&	$Q_h =\frac{1}{b_Q}\bigl\{ Q_m+ c_Q \bigr\}$	      &	Eq.(\ref{eq:b1c2_q3auto})	   \\
             $c_Q^\times$	&	$Q_{h}^\times = \frac{1}{b_Q}\bigl\{ Q_m+ \frac{c_Q}{3} \bigr\}$	&  Eq.(\ref{eq:b1c2_q3cross}) 	  \\
             $c_{\Delta Q}$	&	$c_{\Delta Q} = b_1 \frac{3}{2}[Q_h-Q_{h}^\times]$	      &	Eq.(\ref{eq:cdeltaq}	)	    \\
             $c_{\tau}$	&	$\frac{{ S_{{3}}C_{{12,h}}- C_{{12}}S_{{3,h}} }}{{\displaystyle  \tau_h }}$	      &	\citet{paperI}	    \\
             $c_{\tau}^\times$	&	$\frac{{ S_{{3}}C_{{12,h}}^\times- C_{{12}}S_{{3,h}}^\times }}{{\displaystyle  \tau_h^\times }}$	      &	Eq.(\ref{eq:c2taucross})	    \\
            \hline
         \end{tabular}
      \end{table}

\begin{figure*}
           \centerline{ \includegraphics[width=130mm,angle=270]{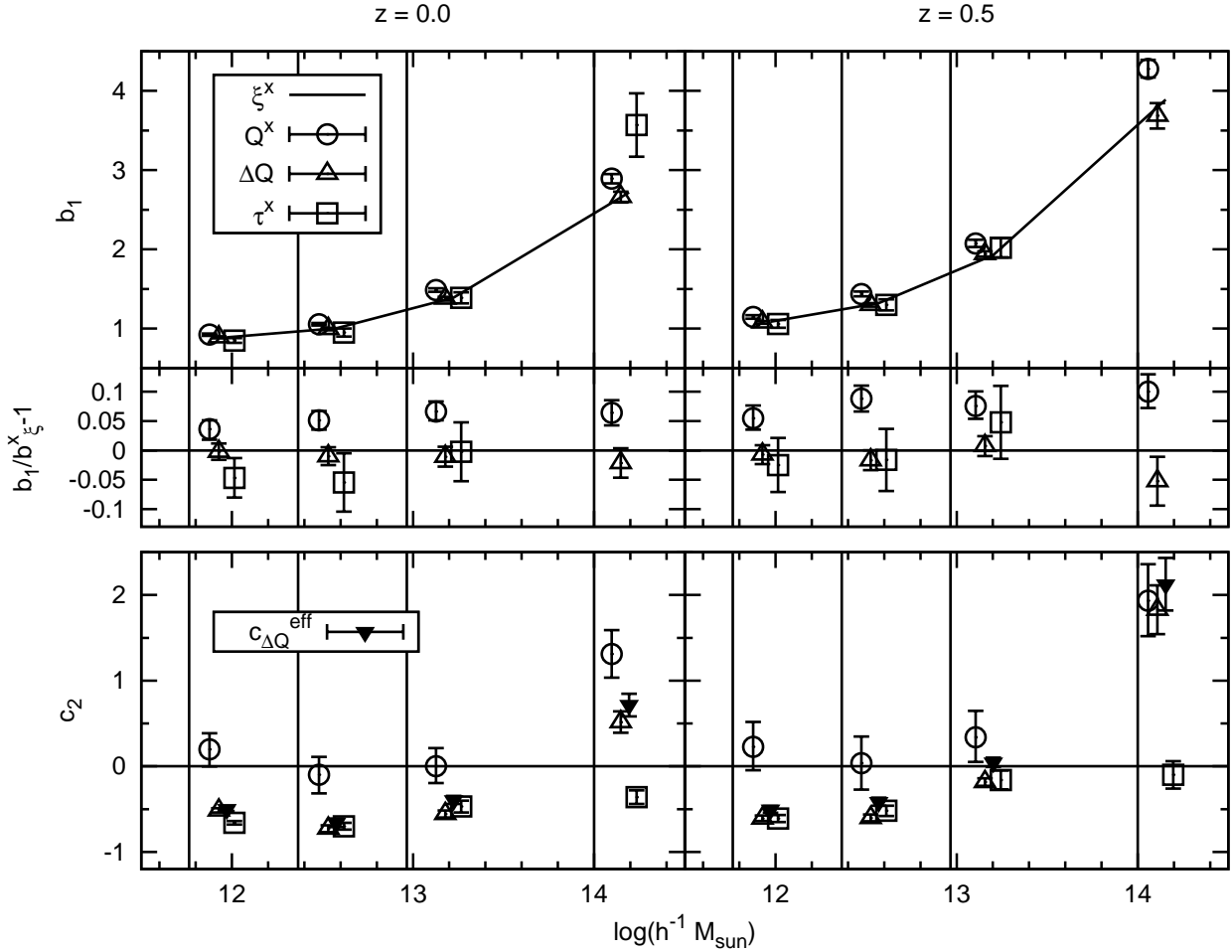} }
          	\caption{\small  {\it Top:} Summary of the linear bias measurements presented in this work.
          	Results are shown for the redshifts $z=0.0$  and $z=0.5$ (left and right respectively) 
          	versus the mean halo mass of each mass sample M0-M3, defined in Table 
          	\ref{table:halo_masses} (lower and upper limits of the mass samples are 
          	marked by vertical grey dashed lines. Symbols are slightly shifted along the the mass axis
          	for clarity). The smaller panel shows the relative deviation of each
          	estimator with respect to the linear bias from the two-point cross-correlation
          	($b_\xi^\times$, solid lines).
          	The estimators $b_{Q}^\times$, $b_{\Delta Q}$ and $b_\tau^\times$
          	(from equations (\ref{eq:b1c2_q3cross}), (\ref{eq:bdQ}) and (\ref{eq:b1taucross}) respectively) are
          	displayed respectively by open circles, open triangles and open squares.
          	For measuring $b_{Q}^\times$ we assumed a local bias model, i.e. $g_2=0$. 
          	{\it Bottom:} Summary of the second-order bias measurements $c_{Q}^\times$, $c_{\Delta Q}$ and $c_\tau$
          	from the same bias estimation methods, as used for the linear bias in the top 
          	panel (equation (\ref{eq:b1c2_q3cross})), (\ref{eq:deltaqcg}) and (\ref{eq:c2taucross}) respectively).
          	The $c_2^{\mathrm{eff}}$ (equation (\ref{eq:c2eff})), obtained by combining $c_2$ and $g_2$
          	from the $\Delta Q$ method, is represented with filled triangles.
          	For estimations based on $Q$ we used triangles with $(r_{12}, r_{13}) = (36,72) \ 
          	h^{-1}$Mpc configurations. Estimations using the $\tau$ method are 
          	fits between $26 \le R \le 40 \ h^{-1}$Mpc. Error bars denote $1\sigma$ uncertainties.
          	}
          	\label{fig:b1c2_comparison}
          	\end{figure*}

          In this section, we aim at comparing the bias estimations coming from these three methods.
          We therefore present the different linear and quadratic bias estimations for the four mass samples
          M0-M3 at the redshifts $0.0$ and $0.5$ versus the mean halo mass in each sample
          in Fig. \ref{fig:b1c2_comparison}. Our bias measurements based on thee-point correlations are done using triangles
          with  $r_{12}=r_{13}/2=36 \ h^{-1}$Mpc. The $\tau^\times$ estimations are based on fits
          of $b^\times_\tau$ and $c^\times_\tau$ between $26 \le R \le 40 \ h^{-1}$Mpc using
          $r_{12}=2R$ configurations.
          
          The linear bias estimations from the different methods are presented in the upper panel of
          Fig. \ref{fig:b1c2_comparison}. We compare these estimations to reference measurements from the
          two-point cross-correlation, defined in equation (\ref{eq:bxicross}), which we consider as
          reliable (see Section \ref{sec:second_order_bias}). The relative deviations to this
          reference linear bias are shown in the central panel.

          We find that the estimator $b_Q^\times$, which neglects the non-local bias, 
          overestimates the linear bias by $5$-$10$\%. The fact that we found a 
          stronger overestimation for $b_Q$ ($10$-$30$\%) in \citet{paperI}
          can be attributed to the lower impact of non-local contributions to the three-point
          cross-correlation compared to the corresponding auto-correlation, as 
          discussed in the previous section.
          
          The linear bias parameters from $\Delta Q$ is in excellent agreement
          with the reference for all mass ranges and at both redshifts.  
          Deviations are in the range of the $1\sigma$ of $b_{\Delta Q}$, while the latter roughly 
          correspond to $1$\% of the amplitude. For the mass sample M3 deviations become
          slightly larger and more significant.
          We find this agreement also for smaller triangle scales, as we demonstrated for $z=0.5$
          in Fig. \ref{fig:b1c2gamma2} and discussed in the previous section.          
          
          Our linear bias measurements from the $\tau_\times$ estimator differs by less 
          than $5$\% from the reference linear bias for the mass samples M0-M2 at both
          redshifts. Such deviation are comparable with the $1\sigma$ errors and are
          therefore not significant. At the highest mass sample M3 the linear bias derived
          from $\tau^{\times}$ deviates significantly from the reference. We found a similar
          behaviour in \citet{paperI} for the $\tau$ estimator. 
          If this deviation was due to exclusion effect (leading to a wrong shot noise correction)
          then we would expect the $b_\tau^\times$
          estimator to be only weakly affected, which is not the case.
          An alternative explanation for these strong deviations can be the presence of
          velocity bias, which possibly cannot be neglected in this mass regime as we
          argued in Section \ref{sec:btau}. 
          
          The lower panel of Fig. \ref{fig:b1c2_comparison} shows how the methods compare in terms of estimating
          the second-order bias parameter $c_2$. Moreover, given that $c_\tau$ only estimates the effective
          second-order bias we also show the $c_2^{\mathrm{eff}}$ (green triangles) computed from
          $c_2$ and $g_2$ (equation (\ref{eq:c2eff})) given by the $\Delta Q$ method. At redshift $z=0.0$ we can
          see that the contribution of the non-local bias to the $c_2^{\mathrm{eff}}$ is very small which,
          given the error on the $c_2^{\mathrm{eff}}$ measured from $c_\tau$, does not allow to separate
          the two measurements. At least, we see that for low mass bins the two effective second-order bias
          values agree within the errors. Hence, we can conclude that at redshift $z=0.0$ the effective
          second-order bias is a good approximation of the second-order bias. Regarding low mass bins at
          redshift $z=0.5$ one can see that, even if the effective $c_2$ from $\Delta Q$ differs significantly
          from $c_2$, it is difficult to say which one is in better agreement with $c_\tau$. The latter seems to be
          in better agreement with $c_2$ than the expected $c_{2}^{\mathrm{eff}}$ coming from the three-point
          correlations. In case of the high mass bin, we observe again at both redshifts that the
          $c_\tau$ estimator delivers unreasonable results. We conclude that the $c_{\tau}$ estimator
          can be used to estimate the second-order local bias parameter as long as the considered 
          tracer are not too massive ($\gtrsim 10^{14} h^{-1}M_\odot$). Finally, we can say that the
          second-order bias, estimated from the  three-point cross-correlation when the non-local bias
          is neglected leads to significant departure from the one obtained from $\Delta Q_{cg}$.  

\section{Summary and Conclusion}\label{sec:conclusion}

 We studied linear, quadratic and, for the first time in configuration space, non-local bias
of halo clustering with respect to the clustering of the dark matter field. 
We therefore employed various second- and third-order statistics of halo and matter density fluctuations
in the MICE-GC simulation.
Our goal was to find if the overestimation of the linear bias parameter by the
three-point auto-correlation, which we found previously in \citet{paperI}
\citep[see also][]{M&G11, Pollack12}, can be attributed to shortcomings of the local
quadratic bias model. Understanding this difference is crucial for breaking the degeneracy
between growth and bias with three-point correlations, which would strongly amplify
the statistical power of large-scale structure surveys. 
To achieve this goal we employ auto- and cross-statistics to disentangle the effects of
linear bias on second- and third-order halo statistics from those originating from
non-linear and non-local bias.

We started our analysis verifying how well the second-order clustering of 
haloes can be described by a linear bias model in Section \ref{sec:second_order_bias}.
Comparing the amplitudes of the variances $\sigma$ to those of two-point correlations
$\xi$ we found the former to be significantly affected by non-linear and
possibly non-local contributions to the bias function. 
However, the halo two-point correlations are well described by a linear local bias model
down to scales of $10 \ h^{-1}$Mpc. For this reason we employed such
measurements from the two-point cross-correlation as an estimator for the linear bias,
used as reference in the subsequent analysis.
Note that we also expect departures from scale independent bias
\citep[see][]{desjacques} for highly biased samples at scales
close the BAO peak \citep[$\gtrsim 80 \ h^{-1}$Mpc,][]{paperIII,paperI}, while 
the large errors at such scales should prevent a strong impact of this effect.

For studying the impact of non-linear and non-local bias on the three-point 
correlation we compared in Section \ref{sec:3pc_bias} bias measurements from
the (reduced) three-point halo-matter cross-correlation to those from the
auto-correlation from our previous study,
using the local quadratic model. We found the linear bias from the 
cross-correlation to be closer to the reference than the linear bias from the 
auto-correlation. This is expected from second-order
perturbation theory, which predicts the three-point cross-correlation to 
be less affected by quadratic local and non-local bias than the corresponding 
auto-correlation (compare equation (\ref{eq:b1c2_q3auto}) and (\ref{eq:b1c2_q3cross})).
However, the three-point cross-correlations delivers, as the corresponding
auto-correlation, linear bias measurements, which lie significantly above the 
reference from the two-point correlation.

To further verify if this overestimation can be attributed to non-linear and 
non-local contributions to the bias model we take advantage of the fact
that three-point auto- and cross-correlations are affected differently by
non-linear and non-local bias, but equally by the linear bias
(see again equation (\ref{eq:b1c2_q3auto}) and (\ref{eq:b1c2_q3cross})).
This property allows for combinations of the auto- and cross-statistics which isolate the linear
from the non-linear and the non-local bias. We find the linear bias, measured by such a 
combination of three-point correlations independently
of quadratic or non-local bias (equation \ref{eq:bdQ}) to be in excellent 
agreement with the reference from the two-point correlation (Fig. \ref{fig:bdQ}).
This finding is a strong indication that non-local terms are indeed the reason for the 
overestimation of linear bias from three-point correlation, when ignoring them 
by assuming a local quadratic bias model. This approach could be used to measure linear bias
by cross-correlating galaxy with lensing maps.
The presence of non-local bias also becomes apparent in our measurements
of non-linear bias contributions (local and non-local) via equation (\ref{eq:deltaqcg}), which
are in good agreement with predictions for the non-local contributions
to the three-point correlation (Fig. \ref{fig:dQnoloc}). Our results therefore constitute 
the first detection of non-local bias in  configuration space and demonstrate the paramount
importance of taking it into account when analysing galaxy surveys.

When the considered scales are two small ($r_{12} \lesssim 30 \ h^{-1}$Mpc), the non-local bias
parameter $\gamma_2$, derived from our measurements, shows a strong
scale-dependence, indicating the presence of higher-order local or non-local terms in the bias 
function. Instead, for scales larger than $36 \ h^{-1}$Mpc we find a linear relation between
the non-local and the linear bias, over the whole mass range, as predicted by the local Lagrangian bias 
model (Fig. \ref{fig:b1-gamma2}). However, the amplitude of this relation
lies significantly below the local Lagrangian prediction.  This is in
agreement with results from \citet{baldauf12} and \citet{Saito14}, but in contradiction with results
from \citet{chan12}, who find the non-local bias to be above the local Lagrangian 
prediction. Whether this latter disagreement comes from the fact that \citet{chan12} analysed
the bispectrum in Fourier space using different simulations is the subject of 
current investigations. An alternative reason for this discrepancy
could be the inaccuracy of the prediction of the non-local contribution to the three-point
correlation of matter, from which we derive the non-local bias
parameter. 

Besides three-point correlations we studied in Section \ref{sec:btau}
bias from one- and two-point  third-order statistics (i.e. the skewness
and the reduced correlator respectively). These two statistics can be
combined into the so-called $\tau$-estimator \citep{bm}, which allows us to
measure the linear and quadratic bias parameters (the latter being local
and non-local) independently from each other. An important difference
to the three-point correlation is that the $\tau$-estimator is an isotropic
quantity. Non-local contributions to the bias function are therefore
absorbed in an effective quadratic bias parameter
(equation (\ref{eq:c2eff})). Hence, the quadratic and non-local bias cannot
be distinguished from each other by the $\tau$-estimator.
Our measurements show that the $\tau$ method delivers a linear bias
estimation which agrees at the $1\sigma$ level with the reference linear
bias from the two-point correlation at scales larger than $20h^{-1}$Mpc,
while the errors on the bias correspond to $\simeq 5$\% of the amplitude
(Fig. \ref{fig:b1c2mznl}).

This result lines up with our findings from the combination of three-point
auto- and cross correlations in the sense that third-order statistics is
able to deliver accurate estimations of the linear bias when the employed
estimator is independent of non-local and possible higher-order
contributions to the bias function.

Interestingly the linear as well as the quadratic bias estimations
derived from the $\tau$ method are failing when measurements
are performed in samples of very massive haloes
($M \gtrsim 10^{14}$M$_{\odot}$). Since this effect appears
in both, the auto- and the cross-statistics it might be attributed
to velocity bias, rather than non-Poissonian shot-noise as we
speculated in \citet{paperI}; further investigations are needed to
understand this effect.

Summarising our various bias estimations in Fig. \ref{fig:b1c2gamma2},
we find an overall variation in the linear bias of $\sim 10$\% with respect to 
the reference from the two-point cross-correlation.
Deviations from the reference are more significant for measurements from the
three-point correlation, which
are based on the quadratic local bias model. The bias estimators, which are independent
from non-local contributions ($\Delta Q$  and $\tau$) do not differ significantly from
the reference for the linear bias, except for $\tau$ in the high mass range, as
we saw previously.

Comparing the quadratic bias from the different estimators we found the effective 
$c_2$, obtained from the $\tau$-estimator, to be close to $c_2$ measured with the combination
of three-point auto- and cross-correlations, which takes into account non-local bias. 
This finding shows that one can reasonably neglect the
non-local bias contribution when using the $\tau$-estimator to measure the quadratic bias
parameter. This neglection is not possible for three-point correlations, as those 
deliver quadratic bias parameters which are significantly higher than 
measurements, which take non-local bias into account.

Our results show that the local quadratic bias model is inadequate to describe 
halo bias in the MICE-GC simulation. Non-local second-order terms need to
be taken into account for accurate measurements of the linear bias with 
three-point correlation function. Two approaches are possible to do so.
Non-local bias can be isolated from linear bias by combining different
third-order statistics (i.e. $\tau$ or $\Delta Q$), or the non-local 
contributions need to be directly modelled. The first approach,
on which we focused in this analysis, might be implemented
in terms of cross-correlations between lensing and galaxy maps. We will
test the second approach in a future analysis, but already provide here
an expression for the non-local contribution to the three-point correlation
in configuration space.
At scales below $30h^{-1}$Mpc we find indications for the presence of
higher-order terms in the bias function (local or non-local). Modelling
the non-local bias as a linear function of the linear bias parameter, as suggested
by the local Lagrangian bias model, therefore appears to be only suitable
at very large scales.

In a future work (Bel, Hoffmann, Gazta\~naga, in preparation) we plan to compare the linear and
second-order bias measurement, obtained in the present analysis to peak-background
split predictions from \citet{paperIII} and also bias parameters obtained by
directly comparing halo- and matter  over densities in the MICE-GC simulation. These
various bias estimations can be used to verify the universal relation between linear
and non-linear bias parameters, which we presented in \citet{paperIII}.



\FloatBarrier
\section*{Acknowledgements}

Funding for this project was partially provided by the Spanish Ministerio de Ciencia e Innovacion (MICINN), project AYA2009-13936,
Consolider-Ingenio CSD2007- 00060, European Commission Marie Curie Initial Training Network CosmoComp (PITN-GA-2009-238356) and research
project 2009- SGR-1398  from Generalitat de Catalunya.
JB acknowledges useful discussions with Emiliano Sefusatti and support of the European Research
Council through the Darklight ERC Advanced Research Grant (\#291521).
KH is supported by beca FI from Generalitat de Catalunya and ESP2013-48274-C3-1-P.
He also acknowledges the Centro de Ciencias de Benasque Pedro Pascual where parts of the
analysis were done.
The MICE simulations have been developed by the MICE collaboration at the MareNostrum
supercomputer (BSC-CNS) thanks to grants AECT-2006-2-0011 through AECT-2010-1-0007.
Data products have been stored at the Port d'Informació Científica (PIC).

We thank Martin Crocce, Pablo Fosalba, Francisco Castander for interesting and useful comments.

\bibliographystyle{mnbst}
\bibliography{paper2.bib}

\appendix
\section{Non-local bias: a perturbative approach}\label{sec:nloc_bias}

In this section, the Fourier transform $f({\bf q })$ or $f({\bf k })$  of a
configuration space quantity $f({\bf x })$ or $f({\bf r })$ is
defined as $f({\bf q}) \equiv (2\pi)^{-3}\int f({\bf x })e^{-i {\bf x }\cdot {\bf q }}d^3 {\bf x }$.

We assume that for large smoothing scales matter perturbations are small
such that they are well described by perturbation theory. We further assume that on such scales,
the biasing function is described by the equation (\ref{eq:def_biasmodel}) of this paper, which is 

\begin{equation}
\delta_h({\bf x}) = b_1  \left\lbrace \delta_{m}({\bf x}) + \frac{c_2}{2}(\delta_{m}^2({\bf x}) - \langle \delta_{m}^2 \rangle)  + \frac{\gamma_2}{b_1} \mathcal{ G}_2({\bf x}) \right\rbrace,
\label{nlocbias}
\end{equation}
where 
$$
\mathcal{ G}_2({\bf x})=- \int \beta_{12}\theta_v({\bf q}_1) \theta_v({\bf q}_2) \hat W[q_{12}R]e^{i {\bf q}_{12}\cdot {\bf x} }d^3 {\bf q}_1 d^3 {\bf q}_2
$$
and $\beta_{12}\equiv 1 - \left( \frac{ {\bf q}_1\cdot {\bf q}_2
  }{q_1q_2} \right )^2$, hence $\langle \mathcal{ G}_2 \rangle = 0$. The
velocity field ${\bf v}$ is normalised in such a way that at linear order its divergence
 is given by the Poisson equation $-\theta_v^{(1)}=\delta_m^{(1)}\equiv \delta_L$.
In the present calculation and in all this paper we work with density fields which are convolved with a
Top-hat window function of typical size $R$. In addition, we
assume that perturbations are small and can therefore be described at second order by 

\begin{equation}
\delta_{m}({\bf x})=\delta_L({\bf x}) + \delta_m^{(2)}({\bf x}), 
\label{perturb}
\end{equation}
where $\delta_L({\bf x})$ is the linear contribution to the
fluctuations. And the second-order contribution \citep[see][]{bernardeau02} can be expressed as 

\begin{equation}
\delta_m^{(2)}=\int F_2({\bf q}_1, {\bf q}_2)\delta_L({\bf q}_1) \delta_L({\bf q}_2) \hat W[q_{12}R]e^{i {\bf q}_{12}\cdot {\bf x} }d^3 {\bf q}_1 d^3 {\bf q}_2,
\label{d2}
\end{equation}
where the second-order perturbation theory kernel is defined as
$F_2({\bf q}_1, {\bf q}_2)\equiv 1+\frac{1}{2}\frac{ {\bf q_1} {\bf q_2}}{q_1q_2}\left( \frac{q_2}{q_1}+ \frac{q_1}{q_2} \right ) - g\beta_{12}$.
This kernel shows that at second order, non-linearities arise due to mode-coupling. The factor $g$ is the second-order
growth factor  which reduces to $2/7$ in an Einstein-de Sitter universe or as long as the growth rate
of structures is given by $f=[\Omega_m(z)]^{1/2}$. This function of time, can be identified with the function $B$ in \citet{catelan95}, or it is related to the function $\kappa$ ($\kappa + g =1/2$) introduced by \citet{bouchet92}, and is aslo related to the $\mu$ function ($\mu + 2g=1$) defined in \citet{kam99}. 

In the following,  in order to use lighter notations we define
$\nu\equiv \delta_h/b_1$.  Using equation (\ref{nlocbias}) and expansion (\ref{perturb}) we can
express $\nu$ at second order as

\begin{equation}
\nu=\delta_L + \frac{c_2}{2}\left [ \delta_L^2 - \langle\delta_L^2\rangle\right ] + \nu^{(2)},
\label{expnu}
\end{equation}
where the quantity $\nu^{(2)}$ is arising from an effective mode coupling kernel such that 

\begin{equation}
\nu^{(2)}\equiv \int F_2^{\mathrm{eff}}({\bf q}_1, {\bf q}_2)\delta_L({\bf q}_1) \delta_L({\bf q}_2) \hat W[q_{12}R]e^{i {\bf q}_{12}\cdot {\bf x} }d^3 {\bf q}_1 d^3 {\bf q}_2.
\label{nu2}
\end{equation}
Note that the effective second-order kernel $F_2^{\mathrm{eff}}$ is obtained by substituting
$g$ by $g+\gamma_2/b_1$ in $F_2$. From equation (\ref{expnu}), taken at three positions
$\bf r_1$, $\bf r_2$ and $\bf r_3$, we can express the three-point correlation function of
$\nu$, $\zeta^\nu \equiv \langle\nu({\bf x_1} )\nu({\bf x_2})\nu({\bf x_3})\rangle$ we obtain

\begin{eqnarray}
\zeta^{\nu} & = & \zeta^L + \frac{c_2}{2}\left [ \langle\delta_L^2({\bf r_1})\delta_L( {\bf r_2})\delta_L({\bf r_3})\rangle + perm \right ] \nonumber \\
                  &   & - \frac{c_2}{2}\left [ \xi^L_{23}\langle\delta_L^2\rangle + perm \right ] \nonumber \\
                 &  & + \langle\nu^{(2)}({\bf r_1})\delta_L({\bf r_2})\delta_L({\bf r_3})\rangle + perm \label{zetanu},
\end{eqnarray}
where use the definitions $\xi^L_{ij}\equiv \langle\delta_L({\bf r_i})\delta_L({\bf r_j})\rangle$ and $\zeta^L \equiv \langle\delta_L({\bf x_1} )\delta_L({\bf x_2})\delta_L({\bf x_3})\rangle$ .
Assuming that the linear part of the density field $\delta_L$ is Gaussian, then its three-point correlation
function is null and its four-point expectation value can be expressed as \citep{Fry84a}
$$
\langle\delta_L({\bf r_1})\delta_L({\bf r_2})\delta_L({\bf r_3})\delta_L({\bf r_4})\rangle=\xi^L_{12}\xi^L_{34}+\xi^L_{13}\xi^L_{24} + \xi^L_{14}\xi^L_{23} 
$$
from which we can derive (${\bf r_1}={\bf r_4}$) that
$\langle\delta_L^2({\bf r_1})\delta_L({\bf r_2})\delta_L({\bf r_3})\rangle = \xi_{23}^L\langle\delta_L^2\rangle + 2 
\xi^L_{12}\xi^L_{13}$.
The first term cancels with the third term of equation (\ref{zetanu}).
As a result we obtain that 

\begin{equation}
\zeta^{\nu}=c_2\left [ \xi^L_{12}\xi^L_{13} + perm \right ] + \mathcal J_{123} + perm,
\label{zetanub}
\end{equation}
where $\mathcal J_{123}\equiv \langle\nu^{(2)}({\bf r_1})\delta_L({ \bf r_2})\delta_L({ \bf r_3})\rangle$.
This expression has been calculated by Bel (in preparation) who has shown that taking into account
the convolution with the spherical Top-hat filter is very well approximated by a Legendre expansion at
second order (which turns out to be exact when no smoothing is applied). Hence

\begin{eqnarray}
\mathcal J_{123} & \simeq & 2( 1 - 2 g/3)\xi^L_{12}\xi^L_{13} + A_R(r_{12}, r_{13}) \nonumber \\
 &  & +  B_R(r_{12}, r_{13})L_1(\mu_{23}) \nonumber \\
  &   &   + \left [ C_R(r_{12}, r_{13}) + 4 g D_R(r_{12}, r_{13}) \right ]L_2(\mu_{23}) , \label{jj}
\end{eqnarray}
where $A_R$, $B_R$, and $C_R$ and $D_R$ are obtained from integrals over the linear power spectrum and
$\mu_{ij}\equiv \cos\alpha_{ij}$. In practice Bel (in preparation) has shown that in the large
separation limit these functions can be identified with
the functions $\xi(r_{12})$ ($\xi_{12}^L$ in our notations) and $\phi (r_{12})$ introduced by \citet{bargaz}. It reads,

\begin{eqnarray}
A_R(r_{12}, r_{13}) & \simeq &  0  \label{afunc} \\
B_R(r_{12}, r_{13}) & \simeq &  - \left[ {\xi_{12}^L}^\prime\phi^\prime (r_{13}) +  {\xi_{13}^L}^\prime\phi^\prime (r_{12}) \right ]\label{bfunc} \\
C_R(r_{12}, r_{13}) & \simeq & 0 \label{cfunc} \\
D_R(r_{12}, r_{13}) & \simeq &  \frac{1}{3}\left[ \xi^L_{12}+ 3\frac{\phi^\prime(r_{12})}{r_{12}}\right] \left [  \xi_{13}^L+ 3\frac{\phi^\prime(r_{13})}{r_{13}}\right], \label{dfunc} 
\end{eqnarray}
where $\phi(r)\equiv\int \dif^3 \mathbf{k} \frac{P(k)}{k^2}W^2(kR) \frac{\sin(kr)}{kr}$ and $^\prime\equiv \frac{\dif }{\dif r}$.

In addition, since we saw that $g \rightarrow g + \gamma_1/b_1$ by substituting it in equation (\ref{jj}) one can split $\mathcal J_{123}$ in two contributions  

\begin{equation}
\mathcal J_{123}=\hat\Gamma_{123} + g_2\mathcal K_{123},
\label{trinu}
\end{equation}
where $ \mathcal K_{123}\equiv \frac{2}{3}\left [  \Gamma_{123} - \xi_{12}\xi_{13}\right ]$ is the non-local part which has been introduced in equation (\ref{eq:def_gamma}) and $\hat\Gamma_{123}$ can be expressed as

\begin{eqnarray}
\hat\Gamma_{123} & = &2( 1 - 2 g/3)\xi^L_{12}\xi^L_{13} + A_R(r_{12}, r_{13}) \nonumber \\
 &  & +  B_R(r_{12}, r_{13})L_1(\mu_{23}) \nonumber \\
  &   &   + \left [ C_R(r_{12}, r_{13}) + 4/3 g D_R(r_{12}, r_{13}) \right ]L_2(\mu_{23}) . \label{gam123}
\end{eqnarray}
From equation (\ref{gam123}), we can express the three-point correlation function of the matter field as 

\begin{equation}
Q_m^{PT}=\frac{\hat\Gamma_{123} + \hat\Gamma_{231} + \hat\Gamma_{312} }{\zeta_H^m},
\label{qmatter}
\end{equation}
which is equivalent to the expression given in \citet{bargaz}. By analogy, we can define a non-local three-point function $Q_{nloc}$ as

\begin{equation}
Q_{nloc} \equiv \frac{\mathcal K_{123} + \mathcal K_{231} + \mathcal K_{312} }{\zeta_H^m}.
\label{qnlocdef}
\end{equation}
The definition (\ref{qnlocdef}) can be written in terms of the function $\Gamma_{123}$ and its 
permutations
 
\begin{equation}
Q_{nloc}= \frac{2}{3}\left\{ \frac{\Gamma_{123} + \Gamma_{231} + \Gamma_{312} }{\zeta_H^L}  - 1\right\},
\label{qnloc}
\end{equation}
where $\Gamma_{123}=D_R(r_{12}, r_{13})L_2(\mu_{23})$ which contains only a quadrupole contribution. 
On the other hand, from equation (\ref{zetanu}), we can express the reduced three-point function of haloes
$Q_h = b_1^{-1}\zeta^\nu/\zeta_H^L$ with respect to $\hat\Gamma_{123}$ and $\mathcal K_{123}$


\begin{equation}
b_1Q_h = c_2 + \frac{\hat\Gamma_{123} + perm}{\zeta_H^L} +g_2\frac{\mathcal K_{123} + perm}{\zeta_H^L}.
\end{equation}
Finally, using definitions introduced before, we find that the reduced three-point function of haloes can be
expressed as 

\begin{equation}
Q_h = b_1^{-1}\left \{ Q_m + c_2 + g_2Q_{nloc} \right\}.
\label{qgal}
\end{equation}
In the end, from equation (\ref{qgal}) we can take the limiting cases which correspond to the
skewness and the correlator. In case of the skewness Bel (in preparation) has shown that
only the monopole contributes. As a result $\Gamma_{123}$ and all its permutations are null therefore 

\begin{equation}
S_{3,h} = b_1^{-1}\left \{ S_3 + 3 (c_2 -\frac{2}{3} g_2) \right\}.
\label{s3gal}
\end{equation}
The second limiting case is obtained when $r_{12}=r_{13}$ and $\alpha_{23}=0$, in this case the
permutations $\Gamma_{231}$ and $\Gamma_{312}$ are null so only the $\Gamma_{123}$
is contributing. As a result we get 

\begin{equation}
C_{12,h} = b_1^{-1}\left \{ C_{12} + 2 [ c_2 -\frac{2}{3} g_2 + 2g_2/3D_R(r_{12}, r_{12}) ]\right\},
\end{equation}
however the function $D_R(r_{12},r_{12})$ contributes by less than $5$\%/$2$\%  on scales
greater/below  than $R=20 h^{-1}Mpc$ compared to $C_{12}$. As a result considering that this
term is multiplied by $2/3g_2$ it will contribute in total at most to $2$\% (high mass bin) and to
less than $0.1$\% for lower mass bins and can be therefore safely neglected, it reads

\begin{equation}
C_{12,h} = b_1^{-1}\left \{ C_{12} + 2( c_2 -\frac{2}{3} g_2) \right\}.
\label{c12gal}
\end{equation}
The same kind of approach can be used to generalize the relations \ref{qgal}, \ref{s3gal} and \ref{c12gal} to their corresponding cross-statistics. Note that those relations are valid at the three-level and do not imply the large separation limit approximation.

\end{document}